\documentclass[12pt]{article}
\usepackage[margin=1in]{geometry}
\usepackage{parskip}
\usepackage{setspace}

\usepackage[english]{babel}
\usepackage[dvips]{graphicx}
\usepackage{amssymb, amsmath}
\usepackage{mathtools}
\usepackage{mathrsfs}
\usepackage{cite}
\usepackage{bbold,bbm}
\usepackage{color}
\usepackage[normalem]{ulem}  
\usepackage{booktabs}
\usepackage{placeins}
\usepackage{comment}
\usepackage{array}
\usepackage{float}
\usepackage{multirow}

\usepackage[bookmarksopen, bookmarksnumbered, bookmarksopenlevel=2, colorlinks=true, linkcolor=blue, filecolor=magenta, urlcolor=blue, citecolor=blue]{hyperref}

\numberwithin{equation}{section}


\usepackage[font={small}, labelfont=bf, margin=0pt, width=\textwidth]{caption}
\newcommand{\bbZ}{\mathbb{Z}}

\newcommand{\bB}{\bar{B}}

\def\Im{{\rm Im\,}}
\newcommand{\dd}{\mathrm{d}}

\usepackage{xcolor}

\begin{document}
\bigskip\ 

\begin{center}
{\LARGE\bfseries Warped Numerical Calabi-Yau Metrics}\\[6mm]

Severin L\"ust,$^{a,}$\footnote{\href{mailto:severin@lpthe.jussieu.fr}{severin@lpthe.jussieu.fr}} Fabian Ruehle,$^{b,c,}$\footnote{\href{mailto:f.ruehle@northeastern.edu}{f.ruehle@northeastern.edu}} Simon Schreyer$^{d,}$\footnote{\href{mailto:simon.schreyer@lmu.de}{simon.schreyer@lmu.de}} \\[10mm]
{
	{\it ${}^{\text{a}}$Sorbonne Universit\'e, CNRS, Laboratoire de Physique Th\'eorique et Hautes \'Energies,\\ LPTHE, F-75005 Paris, France}\\[.5em]	
	{\it ${}^{\text{b}}$ Department of Physics \& Department of Mathematics,\\ Northeastern University, Boston, MA 02115}\\[.5em]
	{\it ${}^{\text{c}}$NSF Institute for Artificial Intelligence and Fundamental Interactions}\\[.5em]
	{\it ${}^{\text{d}}$Arnold Sommerfeld Center for Theoretical Physics,\\ Ludwig–Maximilian–University Munich, Theresienstraße 37, 80333 Munich, Germany}\\[.5em]
}

\end{center}
\setcounter{footnote}{0} 
\bigskip\bigskip

\begin{abstract}
\noindent
We compute numerical warped Type IIB flux backgrounds on Calabi-Yau threefolds following the construction of Giddings, Kachru, and Polchinski. Using physics-informed neural networks, we approximate all three ingredients required by the GKP setup: the Ricci-flat Calabi-Yau metric, the harmonic $(2,1)$-forms representing the imaginary self-dual three-form flux, and the warp factor, which solves a sourced Poisson equation on the internal manifold. We apply our pipeline to the Dwork family of quintics for two different flux vacua, one near a conifold point and one away from it, the latter serving as a numerical cross-check. With these tools, we study the singular bulk problem, and find that for our benchmark point close to the conifold, approximately $0.5$ percent of the total Calabi-Yau volume sits in the throat, and the warp factor is an order of magnitude larger as compared to the bulk. We also introduce several improvements to techniques used for numerical studies of CY metrics and quantities derived from them that might be of interest independently of our application. These include an improved point sampling algorithm that produces samples that are more uniform under the Calabi-Yau measure, a feature-engineered spectral network for the metric, multi-step physics-informed training, and a weighted Huber loss tailored to stiff PDEs with  highly non-uniform sources.
\end{abstract}

\clearpage
\tableofcontents
\clearpage

\section{Introduction}

Compactifications of Type IIB string theory on Calabi--Yau (CY) orientifolds with three-form fluxes \cite{Giddings:2001yu} are a cornerstone of string phenomenology. They underlie the construction of warped throats following Klebanov-Strassler \cite{Klebanov:2000hb}, the moduli-stabilization program of KKLT \cite{Kachru:2003aw}, and many subsequent efforts to engineer realistic four-dimensional vacua with hierarchies, low-scale supersymmetry, or de Sitter solutions.\footnote{
For a recent explicit proposal for candidate solutions, see, e.g., \cite{McAllister:2024lnt}. 
Conceptual concerns about the validity of the KKLT construction have recently been raised in, e.g., \cite{Gao:2020xqh,Lust:2022lfc,Bena:2024are}.} A common feature of these constructions is that they require explicit knowledge of geometric quantities, such as the Ricci-flat metric on the CY, harmonic representatives of cohomology classes, and solutions of sourced PDEs in the internal space, for which no closed form expressions are known for any compact CY threefold of phenomenological interest.

Most quantitative analyses therefore rely on a combination of exact topological data and approximations:
the moduli space is studied via periods, fluxes are characterized by their discrete charges and contributions to the tadpole, and the CY metric is typically ignored. While this information is powerful, it obscures the local geometry of the compactification, which is precisely what one needs to address questions of control, such as how warping is distributed across the bulk of the CY.

Numerical methods aimed at filling this gap have made considerable progress
in recent years. Neural network (NN) approaches initiated in~\cite{Anderson:2020hux,Douglas:2020hpv,Ashmore:2019wzb,Jejjala:2020wcc,Larfors:2021pbb,Larfors:2022nep} have been developed and are publicly available; we will use the  \texttt{cymetric} library~\cite{cymetric}. These methods have been refined further, for example by feature engineering spectral networks \cite{Berglund:2022gvm}, and extended to compute derived quantities such as harmonic forms, eigenvalues of the Laplacian, and physical Yukawa couplings \cite{Butbaia:2024tje,Constantin:2024yxh,Butbaia:2024xgj,Constantin:2024yaz,Constantin:2025vyt}.

The neural networks employed in this context are an instance of physics-informed neural networks (PINNs) \cite{raissi2019physicsinformed}, which approximate solutions of PDEs by treating the residual of the equation as a loss function. PINNs have been applied to a wide range of problems beyond CY metrics, including the Navier-Stokes equations \cite{wang2025discoveryunstablesingularities}, the Nirenberg problem \cite{platt2026nonuniquenesssymmetriesnirenbergproblem,Cortes:2026kfx}, $G_2$-structure manifolds \cite{Douglas:2024pmn}, and various other geometric PDEs in mathematical physics \cite{Hirst:2025seh,Hirst:2026cpm}. The picture that emerges is that PINNs are particularly well-suited to problems where the equation is well understood but explicit solutions are out of reach, exactly the situation we encounter in warped flux compactifications.  

Once a CY metric is available numerically, the next step in the GKP setup is to identify the imaginary self-dual three-form flux $G_3$ and the warp factor $\mathrm{e}^{-4A}$ it sources. Both ingredients are essential for any quantitative statement about the four-dimensional effective theory.
For example, warping-induced backreaction can qualitatively affect the resulting effective description, and may become important in regimes relevant for moduli stabilization and de Sitter uplifts \cite{Bena:2018fqc, Lust:2022xoq, Hebecker:2025tui}.

In particular, large hierarchies require a region in which $\mathrm{e}^{-4A}$ becomes large compared to its bulk value, and the size and shape of this region determines whether the geometry admits a controlled Klebanov--Strassler-like throat at all. The \emph{singular bulk problem} of \cite{Gao:2020xqh} (see also \cite{Carta:2019rhx}) sharpens this concern: the same flux that produces the throat also induces nontrivial warping in the bulk, and one expects generic flux choices to drive a sizeable fraction of the CY into a regime where the supergravity description is questionable. To our knowledge, the size of this strongly warped region has not been quantified numerically for any explicit CY threefold, although the question is clearly relevant to the swampland program and to the validity of the KKLT construction. 

In this paper, we take a first concrete step in this direction. We set up a complete numerical pipeline that, given a choice of flux quanta, (i) determines the stabilized values of the complex structure moduli and the axio-dilaton from the periods, (ii) approximates the Ricci-flat CY metric using the $\phi$-network of \cite{Larfors:2022nep}, (iii) learns harmonic $(2,1)$-form representatives needed to express the ISD flux, and (iv) solves the sourced Poisson equation for the warp factor with a PINN. We apply this pipeline to the Dwork family of quintics for two flux vacua: one stabilizing the moduli close to a conifold point, where a KS-like throat is expected, and one stabilizing them near the Landau-Ginzburg point, where the geometry should be smooth and which serves as a numerical cross-check. Along the way, we implement several technical improvements that we expect to be useful more broadly: an improved point sampling (IPS) algorithm building on \cite{Keller:2009vj,ShiffmanZelditch1999}, a feature-engineered spectral network for the metric following~\cite{Berglund:2022gvm}, multi-step physics-informed training in the spirit of \cite{wang2024multi,AmandaStacked}, and a weighted Huber loss that is essential for resolving the strongly warped throat region. As an application, we use the resulting pipeline to estimate the volume fraction of the CY that is strongly warped near a conifold, finding that the throat region accounts for approximately $0.5$ percent of the total volume. Note that our study was performed on a CY which does not have an orientifold action at the point in moduli space we studied. Furthermore, this point was close but not exponentially close to the conifold. We hope to return to a setup with a proper orientifold involution and study how the singular bulk problem  behaves as a function of distance from the orientifold in such full-fledged models in the future.

The paper is organized as follows. In Section~\ref{sec:GKP}, we review the GKP construction and the equations of motion that our numerical pipeline must solve. Section~\ref{sec:Flux} explains how we identify suitable flux backgrounds on the Dwork quintic and presents our two benchmark vacua. In Section~\ref{sec:Numerics} we describe how the CY metric, the harmonic $(2,1)$-forms, and the warp factor are approximated by neural networks. Section~\ref{sec:Improvements} collects the technical improvements over existing techniques. We present our numerical results in Section~\ref{sec:Results}, including consistency checks against topological data and the Weil--Petersson metric, and our analysis of the size of the throat region. We conclude in Section~\ref{sec:Conclusions} with a discussion of the singular bulk problem in our setup and an outlook on extensions to genuine orientifold backgrounds.

The code accompanying this paper is publicly available on GitHub~\cite{warped-metrics}.

\section{Review of the Giddings-Kachru-Polchinski Setup}
\label{sec:GKP}
We consider warped IIB flux solutions as described by GKP~\cite{Giddings:2001yu}.
The ten-dimensional metric takes the form
\begin{equation}
\label{eq:10DMetric}
    \dd s^2_{10} = \mathrm{e}^{2A(y)} \eta_{\mu\nu} \dd x^\mu \dd x^\nu +  \mathrm{e}^{-2A(y)} \tilde g_{ij}(y) \dd y^i\dd y^j \,,
\end{equation}
where $\tilde{g}_{ij}$ is the Ricci-flat metric on a Calabi-Yau (CY) three-fold.
The warp factor $A(y)$ depends only on the internal CY coordinates $y^i$ ($i = 1, \dots, 6$).
We allow for a non-vanishing five-form flux
\begin{equation}
    \tilde F_5 = (1 + \star_{10}) \, \dd \alpha \wedge \dd x^0 \wedge \dd x^1 \wedge \dd x^2 \wedge \dd x^3 \,,
\end{equation}
where $\star_{10}$ denotes the ten-dimensional Hodge star operator and $\alpha = \alpha(y)$.
We also include non-trivial three-form fluxes
\begin{equation}
    G_3 = F_3 - \tau H_3 \,,
\end{equation}
where $F_3$ and $H_3$ take values in $H^3(\bbZ)$ of the internal Calabi-Yau.
In the supersymmetric case we are interested in, the relevant equations of motion are:
\begin{itemize}
\item Ricci-flatness of the Calabi-Yau metric:
\begin{equation}
\label{eq:RicciFlat}
R_{ij} (\tilde g) = 0
\end{equation}
Importantly, this equation is independent of $\tilde F_5$, $F_3$, and $H_3$. This means we can find the Ricci-flat CY metric in a first step without requiring any other input. 
\item Imaginary self-dual fluxes: 
\begin{equation}\label{eq:ISD}
    \star G_3 = \mathrm{i} G_3 
\end{equation}
Here $\star$ denotes the six-dimensional Hodge star operator corresponding to the Ricci-flat Calabi-Yau metric $\tilde g_{ij}$.
Therefore, this equation depends on the Calabi-Yau metric determined in the previous step,
but is independent of the warp factor $A$ and the five-form flux $\tilde F_5$. Hence, we can solve it using the solution from the first step.
\item The warp factor $A$ and $\tilde F_5$ are related by
\begin{equation}
\mathrm{e}^{4A} = \alpha \,,
\end{equation}
so only one of them is an independent degree of freedom. The relevant equation of motion reads
\begin{equation}
    \label{eq:alphainverseeom}
    -\Tilde{\nabla}^2 \left( \text{e}^{-4A} \right) = \frac{G_{ijk}\,\tilde{\bar G}^{ijk}}{12 \Im \tau} + 2 \kappa_{10}^2 T_3 \tilde \rho_3^\text{loc}\,,
\end{equation}
where 
\begin{equation}
    \tilde \rho^\text{loc}_3 = \sum_i N_i \frac{1}{\sqrt{\tilde g}} \delta^{6}(y-y_i)\,
\end{equation}
describes localized D3-charge sources, such as D3- and D7-branes, as well as O3- and O7-planes.
This equation depends on the previously determined Calabi-Yau metric and three-form flux.
\item Throughout, we assume a constant axio-dilaton,
\begin{equation}
\partial_i \tau = 0 \,.
\end{equation}
This requires imposing certain conditions on D7-branes and O7-planes.
If these conditions are violated, a Ricci-flat metric is in general no longer a solution to the ten-dimensional equations of motion.
See GKP or also \cite{Hebecker:2025tui} for more general solutions of Type IIB supergravity.
\end{itemize}
To summarize, we first determine the metric of the underlying Calabi-Yau manifold. The next step is to determine the $G_3$ flux in such a way that it is imaginary self-dual, i.e.~\eqref{eq:ISD}, and that the flux quanta are chosen such that the complex structure moduli are stabilized close to a conifold point, which is the interesting regime for phenomenological reasons. We use these to compute the warp factor.

\section{Finding Flux Backgrounds}
\label{sec:Flux}
\subsection{General Strategy} \label{sec:strategy}
We start by expanding the 3-form fluxes $H_3$ and $F_3$, and the holomorphic $(3,0)$-form $\Omega$ of the CY $X$, in an integral symplectic basis $(\alpha_I,\beta^I)\in H^3_-(X,\mathbb{Z})$. Here, $H^3_-(X,\mathbb{Z})$ denotes the orientifold-odd piece of the middle cohomology of $X$ and $I=0,\dots, h^{2,1}_-$. The three-form fluxes can be expanded as
\begin{equation}
    F_3 = f^I \alpha_I + f_I \beta^I\,, \qquad \qquad H_3 =h^I \alpha_I + h_I \beta^I\,,
\end{equation}
where $f^I,f_I,h^I,h_I\in \mathbb{Z}$. Furthermore,
\begin{equation}
    \Omega = \Pi^I \alpha_I + \Pi_I \beta^I\,, \qquad\qquad D_{i} \Omega = D_{i}\Pi^I \alpha_I + D_{i}\Pi_I \beta^I\,,
\end{equation}
where $D_{i}$ is the covariant derivative with respect to $i$-th complex-structure modulus, and $\Pi$ is the period vector
\begin{equation}
    \Pi^I = Z^I = \int_{\alpha_I} \Omega\,, \qquad \qquad \Pi_I = \mathcal{F}_I = \int_{\beta^I}\Omega\,.
\end{equation}
The quantity $\mathcal{F}_I= \partial_I \mathcal{F}$ is the derivative of the prepotential $\mathcal{F}$ with respect to $Z^I$.

In the following we write the flux quanta and periods as vectors,
\begin{equation}
    \mathrm{F}_3 =
    \begin{pmatrix}
        f^I\\ f_I
    \end{pmatrix}\,, \qquad
    \mathrm{H}_3 =
    \begin{pmatrix}
        h^I\\ h_I
    \end{pmatrix}\,, \qquad
    \Pi =
    \begin{pmatrix}
        \Pi^I\\ \Pi_I
    \end{pmatrix}\,, \qquad
    D_i\Pi =
    \begin{pmatrix}
        D_i\Pi^I\\ D_i\Pi_I
    \end{pmatrix}\,.
\end{equation}
Here $\text{F}_3$ and $\text{H}_3$ denote the flux vectors obtained by integrating the three-form fluxes $F_3$ and $H_3$ over the chosen symplectic basis of three-cycles.

To find imaginary self-dual $G_3$ flux, we solve the F-flatness conditions $DW=0$ for the axio-dilaton and the complex structure moduli \cite{Giddings:2001yu}, where $W$ is the Gukov-Vafa-Witten superpotential
\begin{align}
    W=\int_X \Omega\wedge G_3
\end{align}
and $D$ is the K\"ahler-covariant derivative, $D_a W= \partial_a W +(\partial_a K) W$, with respect to the complex structure moduli K\"ahler potential
\begin{align}
\label{eq:CSKahlerPotential}
    K= -\ln \left[-i \int_X \Omega\wedge\bar\Omega\,\right].
\end{align}
The F-term equation for $\tau$ reads
\begin{equation}
   0 = D_\tau W = \frac{1}{\overline{\tau}-\tau}\int_X \overline{G}_3\wedge \Omega = \Pi^T \eta \left( \text{F}_3 - \overline{\tau} \text{H}_3 \right) \qquad \Leftrightarrow \qquad \tau =\frac{\overline{\Pi}^T\eta\text{F}_3}{\overline{\Pi}^T\eta\text{H}_3}\,,
   \label{eq:Ftau}
\end{equation}
where we used the standard symplectic product $\eta = \begin{pmatrix}
    0 & \mathbb{1}\\
    -\mathbb{1} & 0
\end{pmatrix}$.

The F-terms for the complex structure moduli are 
\begin{equation}
    0 = D_i W = \int_X D_i \Omega \wedge G_3 = D_i \Pi^T \eta \left(\text{F}_3-\tau \text{H}_3\right)\,.
\end{equation}
Using \eqref{eq:Ftau} this can be rewritten as \cite{Plauschinn:2023hjw}
\begin{equation}
    0 = D_i \Pi^T \eta^T \frac{\text{H}_3 \text{F}_3^T- \text{F}_3 \text{H}_3^T}{N_\text{flux}}\, \eta \,\overline{\Pi}\,,
    \label{eq:Fcs}
\end{equation}
where 
\begin{equation}
    N_\text{flux} = \int_X F_3 \wedge H_3 = \text{F}_3^T\eta \text{H}_3
\end{equation}
is the flux contribution to the D3 tadpole.
Solving \eqref{eq:Fcs} for the complex structure moduli in a fixed flux background determines a point in complex structure moduli space. Since we are ultimately interested in studying a warped KS-like throat, we have to find a flux background that stabilizes the moduli sufficiently close to a conifold point. 

In this flux background, we can then continue with the final step of transforming the flux quanta of the symplectic basis into the harmonic basis where $G_3$ is automatically imaginary self-dual (ISD) and the fluxes are of the form $F_3 = \tilde{f}^i \chi_i + \tilde{f}^{\overline{\Omega}} \overline{\Omega}$
where
\begin{align}
    \tilde{f}^i = & -\frac{1}{\int_X\Omega \wedge \overline{\Omega}}\, (g_\text{CS}^{-1})^{i\bar{\jmath}} \int_X F_3 \wedge \overline{D}_{\bar{\jmath}} \,\overline{\Omega} = - \frac{(g_\text{CS}^{-1})^{i\bar{\jmath}}}{\Pi^T \eta \overline{\Pi}} \text{F}_3^T \, \eta\, \overline{D}_{\bar{\jmath}} \,\overline{\Pi}\,, \\
    \tilde{f}^{\overline{\Omega}} = & \frac{1}{\int_X\Omega \wedge \overline{\Omega}}\, \int_X F_3 \wedge \Omega = \frac{\text{F}_3^T \, \eta \, \Pi}{\Pi^T \, \eta \,\overline{\Pi}}\,,
\end{align}
with
\begin{align}
\label{eq:WPMetric}
    (g_\text{CS})_{i\bar{\jmath}} = \partial_i \partial_{\bar{\jmath}} K
\end{align}
the Weil-Petersson metric on the complex structure moduli space calculated from~\eqref{eq:CSKahlerPotential}. Similar expressions hold for $H_3$ by replacing $F_3\to H_3$.

\subsection{Application to the quintic} \label{sec:quintic}

Let us apply the strategy outlined in Sec.~\ref{sec:strategy} to the Dwork quintic, a one-parameter hypersurface in complex projective space $\mathbbm{P}^4$ with homogeneous coordinates $x_A$, $A=0,\ldots,4$, given by the zero locus of its anticanonical section,
\begin{equation}
\label{eq:DworkQuinticHypersurface}
    \sum_{A=0}^4 x_A^5-5\psi\prod_{A=0}^4 x_A=0\,.
\end{equation}
The periods of the quintic are calculated in \cite{Candelas:1990rm} and can be approximated close to the conifold point up to cubic order in $Z$ as (see e.g.~\cite{Huang:2006hq,CaboBizet:2016uzv})\footnote{We used machine precision in the calculations, but only display the first three significant digits below and in what follows.}
\begin{align}
\label{eq:periods}
    \hspace{-0.3cm}\Pi^1 & = -0.356 \mathrm{i} Z - 0.25 \mathrm{i} Z^2 - 0.195 \mathrm{i} Z^3\,,\hspace{-0.3cm}\\
    \hspace{-0.3cm}\Pi^2 &= 6.2 - 7.11 \mathrm{i} + (1.02 - 0.83 \mathrm{i}) Z + (0.57 - 0.428 \mathrm{i}) Z^2 + (0.402 - 0.288 \mathrm{i}) Z^3\,,\hspace{-0.3cm}\\
    \hspace{-0.3cm}\Pi_1 &= 1.07 -0.025 Z - 0.012 Z^2 - 0.006 Z^3 - \frac{\log[Z]}{2 \pi \mathrm{i}} (2 \pi \mathrm{i} \,0.057 Z - 0.25 \mathrm{i} Z^2 - 0.195 \mathrm{i} Z^3),\hspace{-0.3cm}\\
    \hspace{-0.3cm}\Pi_2 & =1.29 \mathrm{i} + 0.151 i Z + 0.078 i Z^2 + 0.052 \mathrm{i} Z^3\,,\hspace{-0.3cm}
\end{align}
where $Z=1-1/\psi^5$.
Using the explicit expression for the period, we solve \eqref{eq:Ftau} and \eqref{eq:Fcs} by scanning over all possible flux quanta in the range $\{-N_\text{max},\dots,N_\text{max}\}$ (we restricted the range to $N_\text{max}=2$ and found several solutions that stabilized the complex structure moduli close to the conifold locus). Since $N_\text{flux}\geq 0$  for ISD fluxes, we can throw away all combination of flux quanta leading to $N_\text{flux}<0$. 

We are not interested in finding all solutions but only some which stabilize $\psi$ not too close to the conifold point to keep numerics under control but close enough such that the Calabi-Yau admits a proper conical region. Throughout the paper, we will study two solutions: one close to the conifold region and one farther away. For the latter, the numerics should be under much better control, and we use it as a crosscheck. We expand the periods up to order 50 in $Z$, which gives good enough approximations given that the other numerical errors are in the percent range.

\paragraph{Solution close to conifold.} The solution close to the conifold point, located at $\psi=1$ in our conventions, is given by 
\begin{align}
    \text{F}_3=(3, 2, -3, -1)\,,\qquad \text{H}_3=(1, 2, 1, 0)\,.
\end{align}
This gives
\begin{align}
    N_\text{flux}=8\,,\qquad \tau=-0.333+3.6\mathrm{i}\,,\qquad \psi=1.035 -0.017 \mathrm{i}\,.
\end{align}

\paragraph{Solution close to LG point.} To cross-check our numerics, we use a solution close to the Landau-Ginzburg point (which is at $\psi=0$ in our conventions). One solution is given by 
\begin{align}
    \text{F}_3=(0, 3, -3, 0)\,,\qquad \text{H}_3=(3, 0, -1, 0)\,.
\end{align}
This gives
\begin{align}
N_\text{flux}=9\,,\qquad\tau=4.4 + 1.11\mathrm{i}\,,\qquad \psi=0.5 +0.01 \mathrm{i}\,.
\end{align}
We summarize all data in Table~\ref{tab:flux_solutions}.

\begin{table}[h!]
\label{tab:fluxvacua}
\centering
\small
\setlength{\tabcolsep}{6pt}
\renewcommand{\arraystretch}{1.2}
\begin{tabular}{@{}lcc@{}}
\toprule
 & \textbf{Solution close to conifold} & \textbf{Solution away from conifold} \\
\midrule

$F_3$ & $(3,2,-3,-1)$ & $(0,3,-3,0)$ \\
$H_3$ & $(1,2,1,0)$ & $(3,0,-1,0)$ \\
$N_{\mathrm{flux}}$ & $8$ & $9$ \\

$\tau$ & $-0.333 + 3.6\, \mathrm{i}$ & $4.4 + 1.11\, \mathrm{i}$ \\
$g_s$ & $0.278$ & $0.898$ \\
$g_sN_\mathrm{flux}$ & $2.22$& $8.08$ \\ 
$\psi$ 
& $1.035 - 0.017\, \mathrm{i}$
& $0.5 + 0.1\, \mathrm{i}$ \\
\midrule
$\int\Omega\wedge\bar\Omega$ 
& $-16.9\, \mathrm{i}$
& $-4.7\, \mathrm{i}$ \\

$\int\chi\wedge\bar\chi$ 
& $4.28\, \mathrm{i}$
& $0.887\, \mathrm{i}$ \\

$g_{\mathrm{WP}}$ 
& $0.2528$
& $0.1888$ \\

\bottomrule
\end{tabular}
\caption{Two flux vacua solutions obtained by expanding the periods up to order $50$. While we only display 3 (4 for $\psi$ and the WP metric) significant digits, we used machine precision in our computations.}
\label{tab:flux_solutions}
\end{table}

\section{Approximating solutions on the Calabi-Yau}
\label{sec:Numerics}
\subsection{Approximating the CY metric}
\label{sec:ApproximatingCYMetric}
To approximate the CY metric, we  follow~\cite{Anderson:2020hux,Larfors:2021pbb,Larfors:2022nep} and use the \texttt{cymetric} package~\cite{cymetric}. To approximate the CY metric, we use the so-called $\phi$-network. This learns a correction to the K\"ahler potential of the ambient space Fubini-Study metric, such that the pullback of the CY is the approximate Ricci-flat metric,
\begin{align}
    K(x_A,\bar X_A)= K_\text{FS} + K_\text{NN}\,, \qquad K_\text{FS}=t \ln \sum_A|x_A|^2\,,\qquad K_\text{NN}=\phi(x_A,\bar x_A)\,.
\end{align}
Here, $t$ is the K\"ahler modulus which fixes the K\"ahler class of the CY metric. From this K\"ahler potential, we can get the Ricci-flat CY metric $\tilde{g}$ by pulling back the ambient space metric to the anticanonical hypersurface $p=0$,
\begin{align}
    \tilde{g}(y)=p^*(g_\text{Amb}(x_A))=p^*(\partial\bar\partial K_\text{FS} + \partial\bar\partial K_\text{NN})\,.
\end{align}
The K\"ahler class of the metric on the CY is fixed by the K\"ahler class in the ambient space (in this paper, we do not study the K\"ahler moduli sector, so we will simply set $t=1$ throughout). The complex structure dependence of the metric enters explicitly through the pullback, and implicitly since we only care about $K_\text{NN}(X_i)|_X$. The function $\phi$ is learned by sampling points $y_i$ on the CY and demanding that $\phi$ solves a second-order Monge-Amp\`ere equation at each point $y_i$. This condition is encoded in the so-called sigma loss or Monge-Amp\`ere loss
\begin{align}
\label{eq:MALoss}
    \mathcal{L}_\text{MA} = |J^3-\kappa\, \Omega\wedge\bar\Omega|^2\,,
\end{align}
where $\kappa$ is an arbitrary normalization constant (see Section~\ref{sec:ApproximatingCYMetric}), $J=\tfrac{\mathrm{i}}{2}\tilde{g}_{i\bar j}\,\dd y^i\wedge \dd\bar y^{\bar j}$ is the K\"ahler form built from the CY metric, and $\Omega$ is the holomorphic (3,0)-form on the CY. As long as $\phi$ is a function (section of degree zero), this will be well-behaved on patch overlaps. We ensure this by adding another loss term that imposes that $\phi$ be patch-independent. Another possibility is to modify the input of the NN such that it is of scaling degree 0 (and hence any function of it will be as well). We explain this in Section~\ref{sec:FeatureEngineeredSpectralNN}.

\subsection{Approximating the (2,1)-forms}
\label{sec:21Forms}
In this section, we summarize how to compute the (2,1)-forms and the complex structure Weil-Petersson metric following~\cite{Candelas:1987se,Candelas:1989bb,Keller:2009vj}.

In order to calculate the $(2,1)$-forms $\chi_I$, we need to understand how $\Omega$ varies under a change of the complex structure.
We denote the complex structure moduli by $Z^I$ and the holomorphic coordinates on the Calabi-Yau manifold $X$ by $y^a=y^a(Z^I)$. 
The $\chi_I$ are defined by the relation $D_I \Omega = \partial_I \Omega -K_I \Omega=\chi_I$.
Hence we are interested in the $(2,1)$-component of $\partial_I\Omega$, since $K_I\Omega$ is of type $(3,0)$).
The variation of $\Omega$ is given by
\begin{equation}
\label{eq:OmegaDeriv}
    \partial_I\Omega = \frac{1}{3!} \left(\partial_I \Omega_{abc}\right) \dd y^a\wedge\dd y^b\wedge\dd y^c + \frac{1}{2} \Omega_{abc} \,\dd y^a\wedge\dd y^b\wedge \left(\partial_I(\dd y^c)\right)\,.
\end{equation}
The $(2,1)$ component is therefore determined by the variation of $\dd y^a$ under changes of the complex structure. 
To compute this variation, we assume that the Calabi-Yau is realized as a complete intersection of polynomials $p^\alpha$, where $\alpha=1,\dots,N$, in an $N+3$-dimensional ambient space $\mathcal{A}$. We will restrict our discussion here to $\mathcal{A}=\mathbbm{P}^{N+3}$.

The complex structure deformation are then realized by deformations of the polynomials, \
\begin{equation}
    p^\alpha\to p^\alpha+q^\alpha \,.
\end{equation}
The holomorphic coordinates shift under this deformation as
\begin{equation}
    y^a(Z^I+\delta Z^I) \equiv y^a + m^a_{~\alpha}q^\alpha_I \delta Z^I \,,
\end{equation}
where the $m^\mu_{~\alpha}$ are not globally defined, since otherwise they would just be diffeomorphisms. We further find that
\begin{equation}
    \partial_I(\dd y^a) = \dd (m^a_{~\alpha}q^\alpha_I) = \partial_b (m^a_{~\alpha}q^\alpha_I) \dd y^b + (\partial_{\bar{b}}\, m^a_{~\alpha})q^\alpha_I \dd \bar{y}^{\bar{b}}\,,
\end{equation}
where we used that $q^\alpha_I$ only depends on $y$ and not $\bar y$ in the second step. Inserting this into~\eqref{eq:OmegaDeriv} and extracting the (2,1)-piece, we get
\begin{equation}\label{eq:chiOmega}
    \chi_{ab\bar c,I} = \frac{1}{2} \Omega_{abc} \partial_{\bar c}\, m^c_{~\alpha} q^\alpha_I\,,
\end{equation}
Next, we want to find an expression for the $m^a_{~\alpha}$. One way is direct calculation, which gives after some algebra~\cite{Candelas:1987se,Candelas:1989bb}
\begin{equation}
    m^a_{~\alpha} = g^{\bar b a} g_{B\bar A } \frac{\partial \bar x^{\bar A}}{\partial \bar y^{\bar b}} \frac{\partial x^B}{\partial p^\alpha}\,.
\end{equation}
where, as above, $x^A$ are ambient space coordinates, $y^a$ are CY coordinates, $g^{\bar b a}$ is (the inverse of) a metric on the Calabi-Yau,\footnote{The choice of metric does not matter; in particular, this need not be the Ricci-flat metric. Different choices of metrics correspond to different choices of representatives in the same cohomology class.} and $g_{B\bar A}$ is the ambient space Fubini-Study metric.

A second way of calculating $m^a_{~\alpha}$ is by relating it to the extrinsic curvature $K$ on $X$, which is related to $m^a_{~\alpha}$ as 
\begin{equation}
\label{eq:Xdm}
    K_{\bar a~\alpha}^{~b} = \partial_{\bar a} m^{b}_{~\alpha}\,.
\end{equation}
The extrinsic curvature can be written as
\begin{equation}\label{eq:extrinsiccurvature}
   K_{\bar a~I}^{~b} = K_{\bar a~\alpha}^{~b}\, q^\alpha_I = - H_{\alpha\bar\beta} \, g^{\bar b b } \, \frac{\partial\bar x^{\bar A}}{\partial \bar y^{\bar b}} \frac{\partial \bar x^{\bar B}}{\partial\bar y^{\bar a}}  \left(\frac{\partial^2 \bar p^{\bar \beta}}{\partial\bar x^{\bar A}\partial\bar x^{\bar B}} - \hat{\Gamma}^{\bar C}_{\bar A\bar B} \frac{\partial \bar p ^{\bar \beta}}{\partial\bar x^{\bar C}}\right)q^\alpha_I\,,
\end{equation}
where
\begin{equation}
\label{eq:DefinitionH}
    H^{\alpha\bar \beta} = g^{\bB A} \frac{\partial p^\alpha}{\partial x^A} \frac{\partial\bar p ^{\bar \beta}}{\partial \bar x^{\bB}}\,,
\end{equation}
and $H_{\alpha\bar \beta }$ is the inverse of $H^{\alpha\bar \beta}$, i.e.~$H_{\alpha\bar \beta} = \left(H^{-1}\right)_{\alpha\bar \beta}$. The second term in \eqref{eq:extrinsiccurvature} involving the ambient space Christoffel symbols drops out for the ambient space FS metric \cite{Candelas:1987se}. For $g^{\bar b a}$, we choose the pullback of the ambient space Fubini-Study metric.
The resulting $(2,1)$-form $\chi_a$ is in a fixed cohomology class, but it is not harmonic with respect to the Ricci-flat Calabi-Yau metric.
For $\chi_I$ to be harmonic with respect to a given metric $g_{a\bar b}$, it has to fulfill 
\begin{align}
    \Delta_{\bar\partial}\chi_I
    =(\bar \partial\bar\partial^\dagger+\bar\partial^\dagger\bar\partial )\chi_I
    = (-\bar\partial\star\partial\star-\star\partial\star\bar\partial)\chi_I =0\,,
\end{align}
which means that $\bar\partial\chi_I=\bar\partial^\dagger\chi_I=0$, or equivalently $\bar\partial\chi_I=\partial\star\chi_I=0$. The first condition is simply
\begin{equation}
    \label{eq:dbar}
    \bar\partial \chi_I =0 \qquad\Leftrightarrow\qquad \partial_{[\bar a} X_{\bar b]~I}^{~c} = 0\,,
\end{equation}
and the second condition is
\begin{equation}
\label{eq:dstar}
    \partial\star\chi_I=0 \qquad\Leftrightarrow\qquad 
    \xi_{[abc]\bar c,I} = \frac13 \bigl ( \xi_{abc\bar c,I} + \xi_{cab\bar c,I} + \xi_{bca\bar c,I}\bigr) = 0\,,
\end{equation}
where we defined
\begin{equation}
\label{eq:xidef}
    \xi_{abc\bar c,I} = \partial_c\left(\Omega_{abd} g^{\bar d d} g_{e\bar c}  K_{\bar d~\alpha}^{~e}q^\alpha_I\right)\,.
\end{equation}
To obtain a harmonic representative, we have to modify $\chi_a$ by adding an exact piece.
This can be obtained by adding a $\Delta m^a_{~\alpha}$ to $m^a_{~\alpha}$, 
\begin{equation}
\label{eq:tildem}
    \widetilde m^a_{~\alpha} = m^a_{~\alpha} + \Delta m^a_{~\alpha} \,,
\end{equation}
such that $\widetilde m^a_{~\alpha}$ is harmonic with respect to the CY metric $\tilde g$ entering in the Hodge-$\star$ operator.

Lastly, let us briefly describe how to compute the CS Weil-Petersson metric numerically following~\cite{Keller:2009vj}. While we do not need it for the warp factor computation, it is closely related to the (2,1)-forms, and we compute it numerically as a cross-check for our methods. The CS moduli space metric is obtained from an infinitesimal variation of the holomorphic $(3,0)$-form, or alternatively from the K\"ahler potential~\eqref{eq:CSKahlerPotential},
\begin{align}
    \Omega(Z^I+\zeta^I)\approx \Omega(Z^I)+\zeta^I\partial_I\Omega(Z^I)\,,
\end{align}
which gives
\begin{align}
\label{eq:WPCS}
    g_{I\bar J}=-\frac{\int_X\partial_I\Omega\wedge\bar\partial_{\bar J}\bar\Omega}{\int_X\Omega\wedge\bar\Omega}+\frac{\int_X\partial_I\Omega\wedge\bar\Omega\int_X\Omega\wedge\bar\partial_{\bar J}\Omega}{\left(\int_X\Omega\wedge\bar\Omega\right)}
\end{align}
For the derivatives $\partial_I\Omega$ with respect to the complex structure moduli, we get (specializing to a single hypersurface and thus dropping the index $\alpha$)
\begin{align}
\begin{split}
    \partial_I\Omega&=-\frac{1}{\frac{\partial p}{\partial x^{n+1}}}\left(\sum_{A=1}^{n+1}\frac{\partial\vartheta^A_I}{\partial x^A}\right)\bigwedge_{a=1}^n dy^a\\
    &\phantom{=~}-\sum_{a,\bar b=1}^n \frac{(-1)^{n-a}}{\frac{\partial p}{\partial x^{n+1}}}\left(\frac{\partial\vartheta^a_I}{\partial\bar x^{\bar b}}+\frac{\partial\bar x^{\overline{n+1}}}{\partial\bar x^{\bar j}}\frac{\partial\vartheta^a_I}{\partial\bar x^{\overline{n+1}}}\right) \dd y^1\wedge\ldots\wedge\widehat{\dd y^a}\wedge\cdots\wedge \dd y^n\wedge \dd\bar y^{\bar b}\,,
\end{split}
\end{align}
where the derivatives in the sum implement the pullback of $\vartheta^a_I$ to the CY $n$-fold $X$ (in the patch $x^0=1$ and after using $p=0$ to write $x^{n+1}=x^{n+1}(y^1,\ldots,y^n)$), and $\widehat{\dd y^a}$ indicates omission of this term in the wedge product. The $\vartheta^A_I$ are defined as
\begin{align}
    \vartheta^A_I=-(H^{-1})^{\bar\alpha\beta}g^{A\bar B}\frac{\partial \bar p_{\bar\beta}}{\partial x_{\bar B}}\partial_I p_\beta\,,
\end{align}
with $H_{\alpha\beta}$ given in~\eqref{eq:DefinitionH} in the ambient space patch $x^0=1$.

To learn the harmonic (2,1)-forms, we need to learn the correction $\Delta m^a_\alpha$ to $\tilde m^a_\alpha$ in~\eqref{eq:tildem}, which is related to the extrinsic curvature~\eqref{eq:Xdm} and leads to the $(2,1)$-forms via~\eqref{eq:chiOmega}.

The correction $\Delta m$ shifts $\xi$ in~\eqref{eq:xidef} to
\begin{align}
    \xi_{abc\bar c,I}\to \xi_{abc\bar c,I}+\Delta \xi_{abc\bar c,I}\,,\qquad \Delta \xi_{abc\bar c,I}=\partial_c(\Omega_{abd}\tilde{g}^{\bar d d}\tilde{g}_{e\bar c}\partial_{\bar d}\Delta m^e_\alpha q^\alpha_I)\,.
\end{align}
For the loss, we demand that $\chi$ be harmonic, meaning that $\xi+\Delta\xi$ solves~\eqref{eq:dstar}. Note that this computation requires one derivative of the NN learning $\Delta m$ and one derivative of the metric, so a total of three derivatives for the metric learning $\phi$. These are computationally costly, so we precompute the pullbacks, metrics, inverse metrics, derivatives, etc. 

\subsection{Approximating the warp factor}
\label{sec:WarpFactorTheory}
We want to train a NN that approximates the warp factor $\mathrm{e}^{-4A}$ determined by the Poisson equation 
\begin{equation}
\label{eq:warp_factor}
    -\Tilde{\nabla}^2 \left( \text{e}^{-4A} \right) = \frac{G_{ijk}\,\bar G^{\widetilde{ijk}}}{12 \Im \tau} + 2 \kappa_{10}^2 T_3 \tilde \rho_3^\text{loc}\,.
\end{equation}
The right-hand-side has two contributions. First, the flux term, which is positive and is given in terms of the flux data of our flux vacua. The second term has to be negative to allow for tadpole cancellation. This contribution comes from the presence of orientifold planes or curved D7-branes.

The pipeline we explore in this paper requires the interplay of several interwoven, rather non-trivial analytic and numerical methods, so we decided to develop this pipeline for the simplest CY family of Dwork quintics to verify correctness and feasibility. While the quintic does allow for orientifold planes~\cite{Carta:2020ohw}, the Dwork family does not. Hence, in order to be able to solve~\eqref{eq:warp_factor} for this toy model, we mimic the orientifold by a smeared, constant negative energy. We will return to a full-fledged quintic example in the future.

 In order to numerically define the right-hand-side of the warp factor equation \eqref{eq:warp_factor}, we look at the flux term and note that 
\begin{align}
    \frac{G_{ijk}\,\bar G^{\widetilde{ijk}}}{6} \star 1 = G \wedge \star \bar G  
    & = |G^{\bar\Omega}|^2 \bar\Omega \wedge \star\Omega + |G^\chi|^2 \chi \wedge \star \bar \chi\label{eq:fluxterm1}\\
    & = \mathrm{i}|G^{\bar\Omega}|^2 \Omega\wedge\bar\Omega - \mathrm{i} |G^\chi|^2\chi\wedge\bar\chi\,,\label{eq:fluxterm2}
\end{align}
where we used the (anti)-self duality property of $\Omega$ and $\chi$ in the second step. Dividing by the volume then determines the scalar flux term in \eqref{eq:warp_factor}.
There are two possibilities of getting the flux term, corresponding~\eqref{eq:fluxterm1} and~\eqref{eq:fluxterm2}, respectively. 

A priori, the normalization of these terms is not fixed: $\Omega$ is a section of a line bundle and hence only fixed up to normalization. However, in our computation of the $G_3$-flux, we chose a specific normalization for $\Omega$ to compute the periods, which differs from the normalization choice of \texttt{cymetric}. We thus rescale the quantities in~\texttt{cymetric} such that they match the period results stated in Table \ref{tab:flux_solutions}. If our numerical computations (in particular the WP metric) were exact, there would be a single numerical constant relating the CY volume calculated from the WP metric and the CY volume calculated from the holomorphic top form. Due to numerical fluctuations, the factors differ slightly, and we match them separately to the period computation to make sure that we are working numerically with the correct value of the flux contribution to the tadpole.
Let us describe the two methods in more detail:
\begin{itemize}
    \item \textbf{Method 1.} Use the (anti)-self duality property of $\Omega$ and $\chi$, i.e., \eqref{eq:fluxterm2}. With this, we can use the results for $\chi\wedge\bar\chi$ and $\Omega \wedge \bar \Omega$ from the Keller-Lukic method \cite{Keller:2009vj} to calculate the WP metric. This requires a metric on the CY, see~\eqref{eq:xidef}. While this need not be the Ricci-flat one, we used the two-step CY metric in this paper, see Section~\ref{sec:CYMultistep}. This method bypasses the learned harmonic (2,1) forms. The advantage is that one does not need to explicitly make use of inverse CY metrics when writing out the Hodge star operator, and that numerical errors in the NN that approximates the harmonic (2,1)-forms do not propagate to the warp factor computation.  
    \item \textbf{Method 2.} Explicitly evaluate the Hodge star operator~\eqref{eq:fluxterm1} in terms of the two-step CY metric, and use the learned harmonic $(2,1)$-forms.
\end{itemize}

In the paper, we train a NN for the warp factor for each methods and compare the resulting warp factor distributions. We learn the warp factor $A$ with a PINN, modeled by a fully connected NN. Since the full warp factor in~\eqref{eq:10DMetric} is $\exp(\pm 2A)$, it is positive by construction. The computation of the second derivatives are analogous to how we computed them to get the CY metric from the $\phi$ NN: We compute two derivatives of the NN with respect to its (ambient space) input coordinates and pull the result back using the quintic hypersurface equation. As a training objective, we minimize the mean absolute error of the residual of the Poisson equation~\eqref{eq:warp_factor}.

\section{Improvements of Existing Techniques}
\label{sec:Improvements}
\subsection{Improved Point Sampling}
For our numerical computations, we require points on the CY threefold $X$, i.e., zeros of~\eqref{eq:DworkQuinticHypersurface}. For training the NNs to satisfy the Monge-Amp\`ere, the harmonic (2,1)-form and the warp factor equation point-wise, this would be enough. However, if we want to compute integrals (and to make sure that we probe all regions of the CY), we need to know the distribution of the points. The \texttt{cymetric} package implements a point sampling method that generates points on the CY that are uniformly distributed under the pullback of the ambient space FS metric to the CY. To generate this sample, the package uses a theorem by Shiffman and Zelditch~\cite{ShiffmanZelditch1999}. The theorem establishes that the zeros of random sections are distributed uniformly according to the Fubini–Study measure constructed from
\begin{align}
\label{eq:KFS}
K_\text{FS}=\log\sum_{\alpha=0}^N|s_\alpha|^2=\log\sum_{\alpha=0}^N s^\dagger \lambda^{-1} s
\end{align} where $s_\alpha$ are line bundle sections and $\lambda^{-1}=\mathbbm{1}$ for the standard FS metric. We will use $s_\alpha\in\Gamma(\mathcal{O}_{\mathbbm{P}^4}(1))$, i.e., monomials $s_\alpha=x_\alpha$ in the homogeneous ambient space coordinates.\footnote{Note that one could also choose higher order polynomials potentially improving the results we present here. We only checked second-order polynomials without noticing any improvement in the results.}
For details we refer to \cite{Larfors:2022nep}.

Ultimately, we perform integrals with respect to the CY metric and not the pullback of the FS metric. Since 
\begin{align}
    \Omega\wedge\bar\Omega \propto \sqrt{\det \tilde{g}}\; \dd y^1\wedge\dd y^2\wedge\dd y^3\wedge\dd\bar y^{\bar1}\wedge\dd \bar y^{\bar2}\wedge\dd\bar y^{\bar3}\,,
\end{align}
and we can compute $\Omega$ analytically in a given patch, we know the determinant of the CY metric $\det{\tilde g}$ that enters the measure, and we can use this to re-weight the FS sample to take into account regions where a uniform sample under the FS metric produces an over- or under-sampled region under the CY metric. The FS metric tends to sample the bulk very well but miss small regions with high curvature~\cite{Ahmed:2023cnw}.

The idea of Keller and Lukic in~\cite{Keller:2009vj} is to not just sample all points with the FS metric, but to sample with a collection of FS-like metrics by adapting $\lambda$ in~\eqref{eq:KFS} to produce a sample that is distributed more uniformly under the CY metric. Note that under the CY metric, the mass (or point weight)
\begin{align}
\label{eq:MassFormulaPointSample}
    m(x)=\kappa \left.\frac{\Omega\wedge\bar\Omega}{J^3}\right|_{x}
\end{align}
is equal to 1 for each point $x$ on the CY. Of course, we do not know the CY metric and we need points on the CY to approximate it in the first place. Constructing $J$ from a metric on $X$ that is not the Ricci-flat metric, the mass function can fluctuate strongly point-wise, i.e., $\text{max}(m(x))/\text{min}(m(x))\gg1$, which indicates that some regions are vastly over- and undersampled.
This can lead to larger errors in training the NN and in the numerical Monte-Carlo integration.

As Keller-Lukic point out~\cite{Keller:2009vj}, one can easily find a $\lambda$ such that the associated K\"ahler metric leads to a mass $m(x)=1$ at a fixed point $x$ (of course, it will not be equal to 1 anywhere else, unless this $\lambda$ is giving rise to the actual CY metric, which it is not). So the point sampling algorithm is as follows:
\begin{enumerate}
    \item Sample one region using the FS metric, $g_\text{FS}=g_0$
    \item Find the points $x_\text{min}$ and $x_\text{max}$ that currently have the smallest and largest mass~\eqref{eq:MassFormulaPointSample} under the FS metric
    \item Construct $\lambda_\text{min}$ and from it a new metric $g_1$ such that it has mass $m(x_\text{min})=1$,  sample points with respect to this metric, and add them to the point sample.
    \item Construct $\lambda_\text{max}$ and from it a new metric $g_2$ such that it has mass $m(x_\text{max})=1$ and sample points with respect to this metric, and add them to the point sample.
\end{enumerate}
Steps 2-4 can be repeated multiple times, each time finding the points $x$ that have the smallest and largest mass under the currently used metrics. By construction, this gives a collection of point sets sampled from a collection of FS-type metrics $g_i$ such that the sampled points have masses close to 1 under (at least) one of the metrics used to generate the full point sample, and hence leads to a point sample that is distributed more closely according to the Ricci-flat metric than using just a single matrix $g_0$ for the entire point sample.

To find the metric under which a given point $x$ has mass one, we construct the $\lambda$ in~\eqref{eq:KFS} as
\begin{equation}
\label{eq:Lambdas}
    \lambda_x = \frac{1}{1+\varepsilon} \left(\mathbb{1}+ \varepsilon P_x\right)\,, \qquad \varepsilon = m(x)^{1/n} -1\,,
\end{equation}
where $P_x$ is the projector that projects $x$ onto the ray of sections $s_\alpha$ and $n=3$ for CY threefolds. In our case where $s_\alpha = x_\alpha$, $P_x$ is
\begin{equation}
    (P_x)_{\alpha\bar \beta} = \frac{x_\alpha \bar x_{\bar\beta}}{|x|^2}\,.
\end{equation}
In order to apply the Shiffmann-Zelditch theorem, we need a collection of random sections, where random means that their numerical coefficients are iid Gaussian. For $\lambda^{-1}=\mathbbm{1}$, these are Gaussians with zero mean and unit variance. Once we use different $\lambda$, we adjust the variance based on the Cholesky decomposition of $\lambda$, which is $\lambda=L^\dagger L$ since $\lambda$ needs to be a positive definite Hermitian matrix such that the logarithm in~\eqref{eq:KFS} gives a real K\"ahler potential.

There are several variations to this sampling method. One is to implement rejection sampling where one rejects a point sampled with a metric $g_i(\lambda_x)$ if there already exists another metric under which this point would have a mass function closer to one. Another variation to reduce cross-correlation between regions is to not use points $x$ that have already been sampled to determine the new regions based on $\min(m(x))$ and $\max(m(x))$, but to determine these regions from an independent set of points that are not included in the sample.

\begin{figure}
    \centering
    \includegraphics[width=0.31\textwidth]{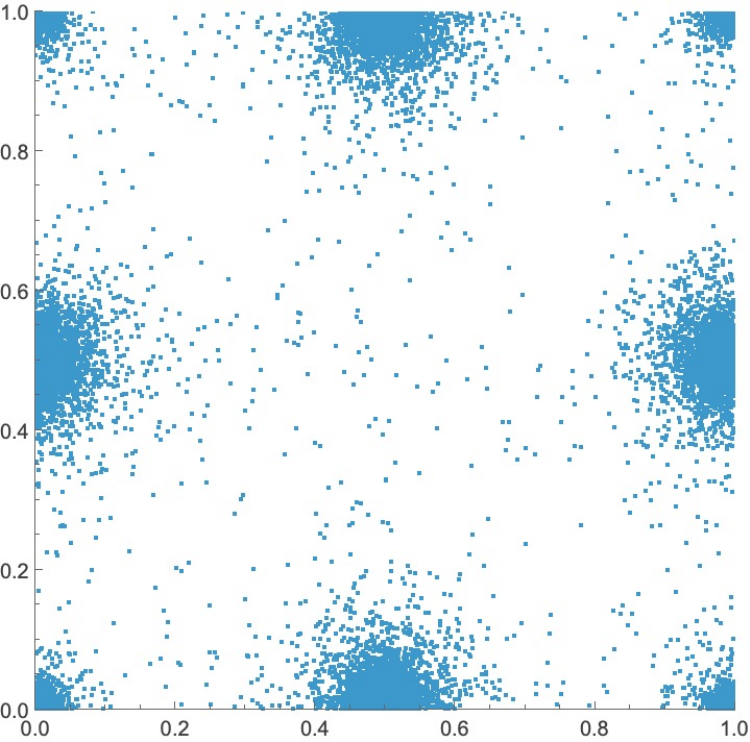}
    \includegraphics[width=0.31\textwidth]{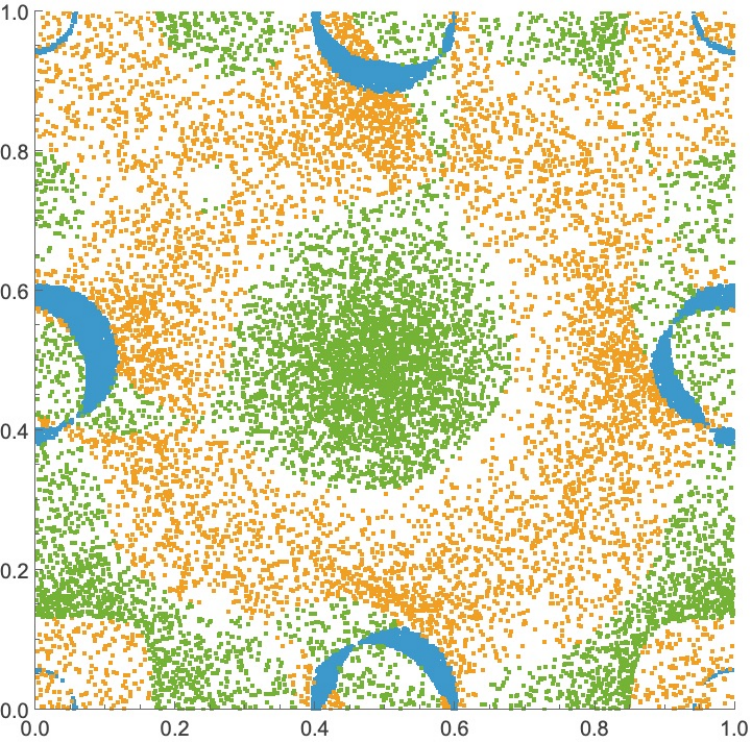}
    \includegraphics[width=0.31\textwidth]{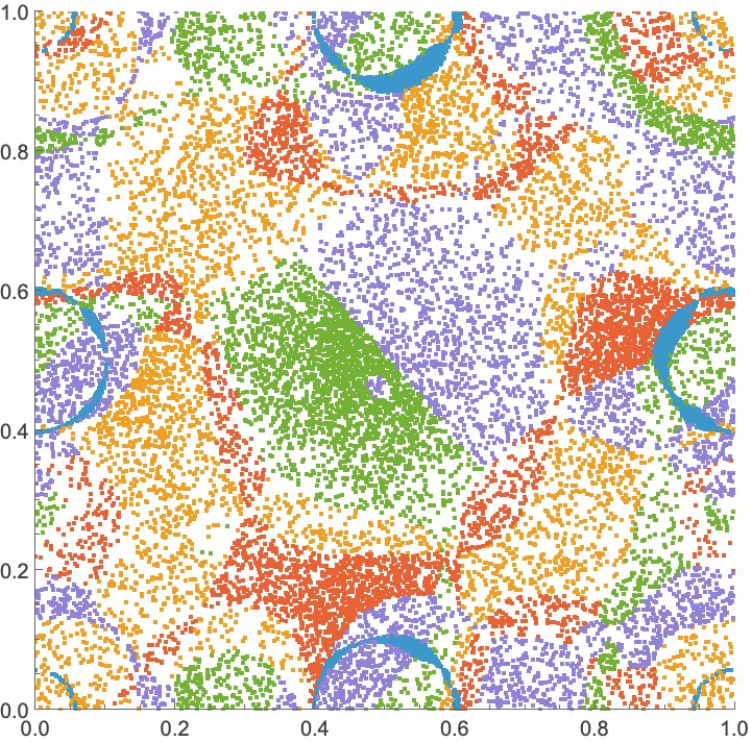}
    \caption{Distribution of points on a square torus lattice when sampled with 1,3, and 5 different metrics using point rejection.}
    \label{fig:T2IPSExample}
\end{figure}

To illustrate the effect of the improved point sampling, we use the one-parameter family of cubics in $\mathbbm{P}^2$, 
\begin{align}
    x_0^3+x_1^3+x_2^3-3\psi x_0x_1x_2=0\,,
\end{align}
which is the CY one-fold analog of the Dwork family of quintics. In this case, the CY is a torus, and the Ricci-flat CY metric is just the flat metric on $\mathbbm{C}/\Lambda$. For illustration purposes, we choose $\psi$ such that $\Lambda=\mathbbm{Z}+\mathrm{i} \mathbbm{Z}$ is the square lattice as in~\cite{Keller:2009vj}. A uniform sample under the CY metric would thus produce a set of points that is equally distributed within the square. This is demonstrated for different number of regions in Figure~\ref{fig:T2IPSExample}.

\subsection{Feature-engineering a spectral NN}
\label{sec:FeatureEngineeredSpectralNN}
The idea of a spectral NN was put forward in~\cite{Berglund:2022gvm}: If the input to the $\phi$ NN of~\cite{Larfors:2022nep} is invariant under the projective scalings of the ambient space, it is automatically well-behaved under transition maps on the manifold. Moreover, since the NN is now a globally defined real function, it cannot change the K\"ahler class, and we can also drop the volume loss (although keeping or dropping it does not have a big impact on performance). This can be arranged easily: instead of using the homogeneous coordinates $x_A$, $A=0,\ldots,4$ (or rather, their real and imaginary parts, since NNs are typically real) as inputs, we provide the inputs 
\begin{align}
f_{A\bar B}= \frac{x_A \bar x_{\bar B}}{|x|^2}\,,
\end{align}
which, by construction, have scaling degree 0. This allows us to drop the transition loss and leads to better results on near-singular CYs~\cite{Berglund:2022gvm}. The obvious disadvantage is that the input (and hence the number of NN parameters) grows substantially. For the quintic, the NN would usually get 10 inputs (the real and imaginary parts of the 5 homogeneous ambient space coordinates), while the feature-engineered NN has $50$ inputs (the real and imaginary part of the $5\times 5$ features $f_{A\bar B}$).\footnote{The number of independent features is a bit lower: $f_{A\bar A}$ has no imaginary parts, and $f_{A\bar B}$ and $f_{B\bar A}$ have the same real part and their imaginary part differs only by a sign, which can be absorbed in the NN parameters.} Since PINNs can often be chosen rather small and training is fast in either case, we will typically use the spectral NN to approximate the CY metric.

\subsection{Learning the CY metric via multi-step training}
\label{sec:CYMultistep}
As discussed in Section~\ref{sec:ApproximatingCYMetric}, we approximate the CY metric by learning a correction to the ambient space FS K\"ahler potential. Instead of capturing all corrections with a single NN, it has been proposed~\cite{wang2024multi,AmandaStacked} that one uses multiple PINNs, each refining the result of the previous one. Concretely, in our case, this means that we would have multiple $\phi$-NNs that approximate the CY metric as the pullback of
\begin{align}
    \tilde{g}=\partial\bar\partial \left[K_\text{FS} + a_1\phi_1 + a_2\phi_2 +\ldots\right]\,,
\end{align}
where the $a_i$ are real parameters controlling the strengths of the corrections. One then first trains $\phi_1$ (we use $a_1=1$) until the loss does not improve significantly anymore. At that stage, the weights of $\phi_1$ are frozen and we add $\phi_2$ in. This is then trained again until no significant improvement of the loss is achieved. One could initialize the $a_i$ to be of the order of the residuals of the PDE we are trying to solve, but we found that even upon setting all $a_i=1$, the NNs quickly learn to produce output of the right magnitude, i.e., such that they only slightly correct the previous NNs.

\begin{figure}[t]
    \centering 
    \includegraphics[width=0.7\textwidth]{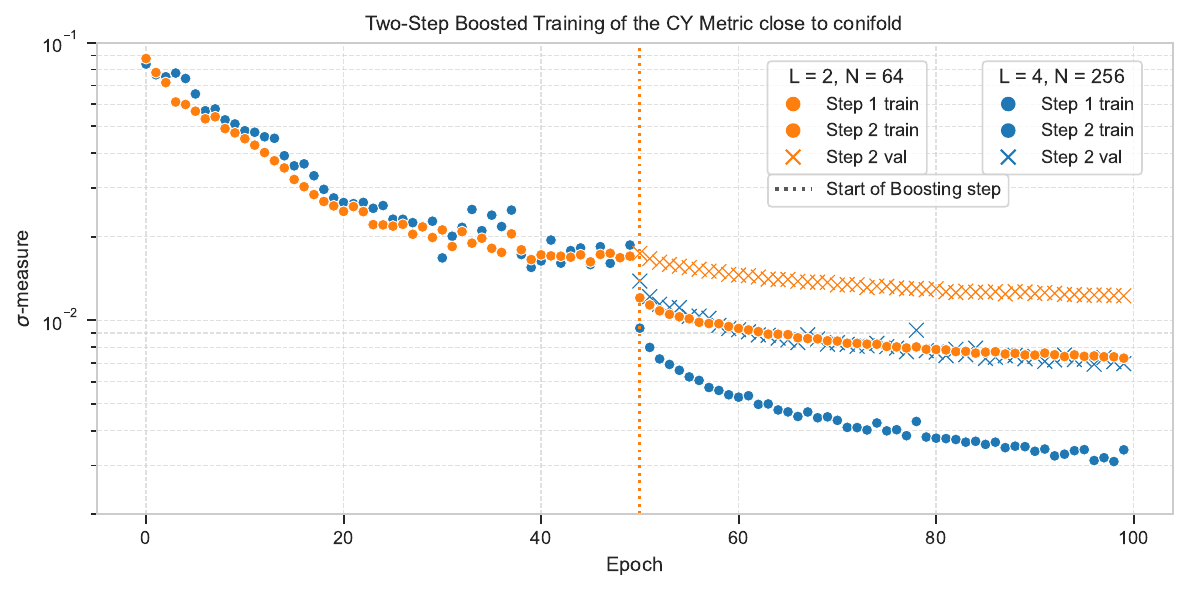}~~
    \caption{Two-step training of the CY metric for two different model sizes (2 hidden layers with 64 neurons, and 4 hidden layers with 256 neurons).}
    \label{fig:training_twostep}
\end{figure}

We illustrate the effect of a two-step PINN for two different NN architectures in Figure~\ref{fig:training_twostep}.  We trained on 100k points with ADAM optimizer (and the default batch sizes and learning rates of~\texttt{cymetric}) and GELU activation function. The first NN was trained for 50 epochs, frozen, and then the second NN was trained for another 50 epochs. As we can see from the plot, adding the second NN introduces some overfitting, but the error still decreases considerably: After stalling around $2\times 10^{-2}$ for about 20 epochs, the test error drops to $2\times 10^{-3}$ after adding the second NN.

\subsection{Using second-order optimizers for NN training}
\begin{figure}[t]
    \centering 
    \includegraphics[width=0.75\textwidth]{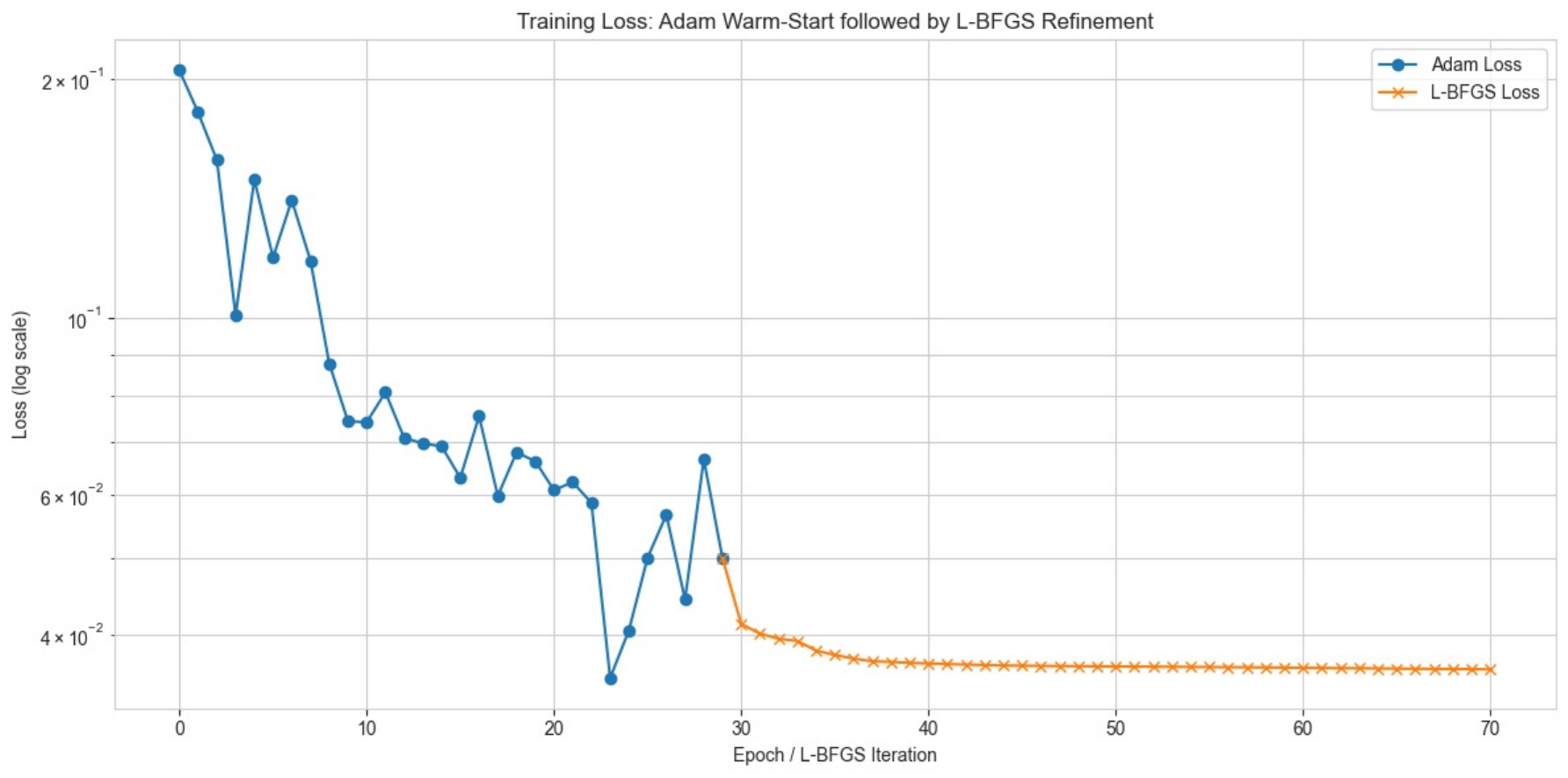} 
    \caption{Training the CY metric $\phi$ NN for 30 epochs with ADAM, followed by 40 epochs with L-BFGS.}
    \label{fig:LBFGSTraining}
\end{figure}

While modern transformer models have trillions of parameters, PINNs can often be chosen much less sophisticated. Indeed, the networks we use here are vanilla feedforward NNs with 2-4 hidden layers with tens or hundreds of neurons per layer, resulting in models with thousands or at most tens of thousands of parameters. This means that it is feasible to train PINNs with second-order optimizers such as L-BFGS~\cite{L-BFGS}. A training procedure where a PINN is first trained with ADAM and then with a few steps of L-BFGS was already proposed in~\cite{raissi2019physicsinformed}. We implemented this strategy using \texttt{scipy}'s L-BFGS optimizer, where we trained a small NN (2 hidden layers, 32 neurons each) on 100k points close to the conifold for 30 epochs with ADAM, followed by 40 epochs of L-BFGS training. We show the result in Figure~\ref{fig:LBFGSTraining}. As we can see, using a second order optimizer is not worth it in this case: One epoch (100k points) takes 5 seconds with ADAM and 20 seconds with L-BFGS for this NN. The factor of 4 in compute cost leads to nearly no improvement of the loss, so we decide to train with ADAM for longer instead of switching over to L-BFGS.

\subsection{Weighted Huber loss for the warp factor}
Training with just an MAE loss on the Poisson equation~\eqref{eq:warp_factor} tends to smoothen out extreme outliers. However, we are interested in points in the throat: There are very few of these in our data set (since the sampling method undersamples strongly curved regions), and they are outliers in the sense that warping is very strong in the bulk. This is reflected in a much larger-than-average right hand side in~\eqref{eq:warp_factor}. 

Without taking both effects into account, we will not be able to make the throat region visible numerically. Therefore, we introduce a weighted loss function that up-weighs the loss contribution of the throat region, using a weighted Huber loss,
\begin{align}
\mathcal{L}_\text{warp} = 
\frac{\sum_x w_{\mathrm{gt}}(y)\,
\mathrm{Huber}\big(y, f(x)\big)}
{\sum_x w_{\mathrm{gt}}(y)}\,,
\end{align}
where the Huber loss is defined as
\begin{align}
    \text{Huber}_\delta(y,f(x))=\left\{\begin{array}{ll}
         \frac{1}{2}(y-f(x))^2&\qquad|y-f(x)|\leq\delta  \\[4pt]
         \delta(|y-f(x)|-\frac12\delta)& \qquad\text{else}
    \end{array}\right.\,,
\end{align}
and $y$ is the source at point $x$. In all examples below, we use $\delta =0.3$. For small enough $|y-f(x)|$, it behaves like a mean-squared error loss, and for large $|y-f(x)|$, it behaves like a mean absolute error loss.
The weighting function emphasizes rare but large negative values of $\rho(x)$ while saturating extreme outliers,
\begin{align}
w_{\mathrm{gt}}(y)
= 1 + \alpha \cdot \min(\max(0,-y),\, c)  = 1+ 10 \min\left(\max(0,-y),5\right),
\end{align}
where $\alpha$ is used to emphasize outliers up to value $-c$. The Huber loss shows better and more stable convergence for second derivative problems. As with the other NNs, the model is fully physics-informed: the warp factor is not learned in a supervised fashion, but via enforcing that the residuals of the differential equation~\eqref{eq:warp_factor} vanish point-wise, with a loss adapted for highly non-uniform and stiff source terms.

\section{Results}
\label{sec:Results}
\subsection{Consistency checks} \label{sec:checks}
Besides just seeing the error decreasing during training, we want to test our implementations of the NNs and the point sampling methods. Since the quantities we want to analyze are not known analytically, the idea is to use our implementation to compute derived quantities which are topological and known either analytically or can be computed approximately by other means. A good example is the Euler number of the manifold: It is topological, independent of the moduli (as long as the CY is smooth) and known analytically from an index computation. This means that we can compare the exact result $\chi=-200$ with the numerical result from our two benchmark points. Since it is topological, we can also use either the pullback of the Fubini-Study metric (which is exact up to machine precision), or the approximate CY metric (which involves a NN). This allows us to see differences in the point sampling and check that the NN implementation for the CY metric is well-behaved.

A second quantity that we can use for our comparison is the Weil-Petersson metric~\eqref{eq:WPMetric}, which can be computed in two ways: From the periods by expanding the hypergeometric solution to the Picard-Fuchs equation, and by varying the metric. 

\paragraph{Euler number.}
To compute the Euler number we compute the Chern classes from the Riemann tensor following~\cite{Berglund:2022gvm} and integrate it using the standard Monte-Carlo integration method of the~\texttt{cymetric} package. We find that the result away from the conifold is within less than one percent of the exact result $\chi=-200$, independent of which metric we use (the FS metric, the CY metric, the CY metric with feature engineering, or the two-step CY metric with feature engineering). This is encouraging, since the integral should be topological and hence metric-independent. We take this as a cross-check that away from singularities, our implementation is correct. Close to the conifold singularity, the quality of the Euler number computation deteriorates, with errors now being around ten percent. The error is again mostly independent from the metric we use, and is already high for the FS metric, which is exact up to machine precision. This means that the error comes from the point sample. Numerically, this is not too surprising: close to the conifold, the CY contains regions with strong positive and negative curvature. In the end, the contribution of these regions have to cancel to give the result from the smooth case, but for large values this cancellation is prone to numeric instabilities.

Using a different point sampling method that samples more regions does hurt more than it helps in this case: The regions will be constructed to sample the strongly curved regions more densely. With more points in these regions, there is more room for cancellation errors in the Euler number computation. These numerical instabilities come with either sign and while the error in the mean reported in Table~\ref{tab:ConsistencyChecks} is slightly lower for the IPS sampling methods, the variance is quite high.

\begin{table}[t]
\centering
\small
\setlength{\tabcolsep}{8pt}
\renewcommand{\arraystretch}{1.2}
\begin{tabular}{@{}lccc@{}}
\toprule
Solution 
 & Exact value
 & Standard sampling 
 & IPS (no rejection)\\
\midrule

$\chi$ close to conifold & $-200$
& $-228.2$
& $-187.3$ \\

$\chi$ away from conifold & $-200$
& $-199.9$
& $-200.7$ \\

\midrule

$g_\text{WP}$ close to conifold
& $0.2528$
& $0.2318$
& $0.2316$ \\

$g_\text{WP}$ away from conifold
& $0.1894$
& $0.1884$
& $0.1906$ \\
\bottomrule
\end{tabular}
\caption{Mean Euler number and WP metric (4 significant digits) for the two flux vacua using the FS metric, comparing two different sampling methods. For IPS we sampled five different regions.}
\label{tab:ConsistencyChecks}
\end{table}

We summarize our results for the FS metric in the top rows of Table~\ref{tab:ConsistencyChecks}. 
To smoothen out the numerical effects discussed above, we compute the Euler number for 10 different data sets of 100k points each. We give the mean value of the Euler number at our two benchmark points for standard sampling and for IPS sampling with 5 regions. We find that the result away from the conifold is extremely accurate for either sampling method. Close to the conifold, errors go up to around 10 percent.
For certain datasets we have calculated the Euler number additionally using the CY and the multi-step CY metric using IPS and standard sampling as shown in Table \ref{tab:boosted_euler}. For most examples, this significantly improves the precision of the Euler number. Note that in theory, the Euler number is a topological invariant and hence independent of the metric. The fact that we find different results for different metrics indicates that i) the point sample is not fully representative of the manifold if it is close to developing a conifold singularity and ii) that the single-step CY metric starts to overfit with high-frequency modulations, which are smoothed out if multiple NNs are used in the approximation.

\begin{table}[t]
\centering
\small
\setlength{\tabcolsep}{6pt}
\renewcommand{\arraystretch}{1.15}
\begin{tabular}{ccccc}
\toprule
Dataset & Sampler & FS metric & CY metric & CY metric (two-step) \\
\midrule

\multirow{2}{*}{$0$}
& Standard & $-259$ & $-197.9$ & $-194.2$ \\
& IPS      & $-227.9$ & $-233.1$ & $-234.9$ \\
\addlinespace

\multirow{2}{*}{$6$}
& Standard & $-263.3$ & $-264.2$ & $-203.2$ \\
& IPS      & $-247.2$ & $-246$ & $-210.6$ \\
\addlinespace

\multirow{2}{*}{$8$}
& Standard & $-228$ & $-247.2$ & $-270$ \\
& IPS      & $-229.6$ & $-204.8$ & $-211.6$ \\

\bottomrule
\end{tabular}
\caption{Euler number $\chi$ (4 significant digits) for the solution close to the conifold computed using different metric approximations.
Each dataset contains $10^5$ sampled points.}
\label{tab:boosted_euler}
\end{table}

\paragraph{Weil-Petersson metric.}
The Weil-Petersson metric~\eqref{eq:WPCS} can be written entirely in terms of ambient space quantities and their pullbacks, without reference to the CY metric (the only target space metric that appears is the FS metric of the ambient space). It therefore serves as another good crosscheck for the point sample. The results are in the bottom rows of Table~\ref{tab:ConsistencyChecks}. As for the Euler number, the WP metric approximation is very good away from the conifold (sub-percent error), while its error is around 10 percent close to the conifold.

\subsection{Learning the CY metric}
Learning the CY metric on the Dwork family of quintics has been benchmarked extensively, see~\cite{Larfors:2022nep} for the \texttt{cymetric} package. Using a small NN to approximate the CY metric, we obtain one to two orders of magnitude improvement over using the pullback of the FS metric. The two-step procedure outlined in Section~\ref{sec:CYMultistep} helps obtain better results faster, see Figure~\ref{fig:training_twostep}, and using sections as inputs gives better results close to the conifold~\cite{Berglund:2022gvm}. We use both improvements to approximate the CY metric, which enters in the (2,1)-forms and the warp factor computation.

\subsection{Learning harmonic (2,1)-forms}

\begin{figure}[t]
    \centering 
    \includegraphics[width=0.45\textwidth]{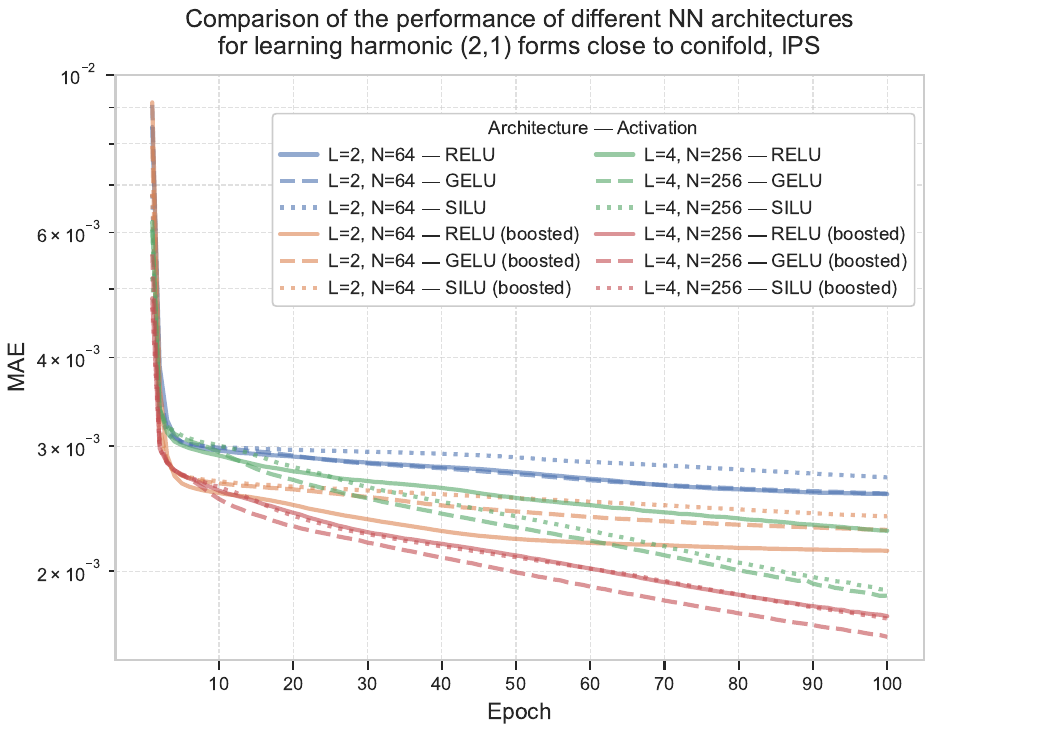}
    \includegraphics[width=0.45\textwidth]{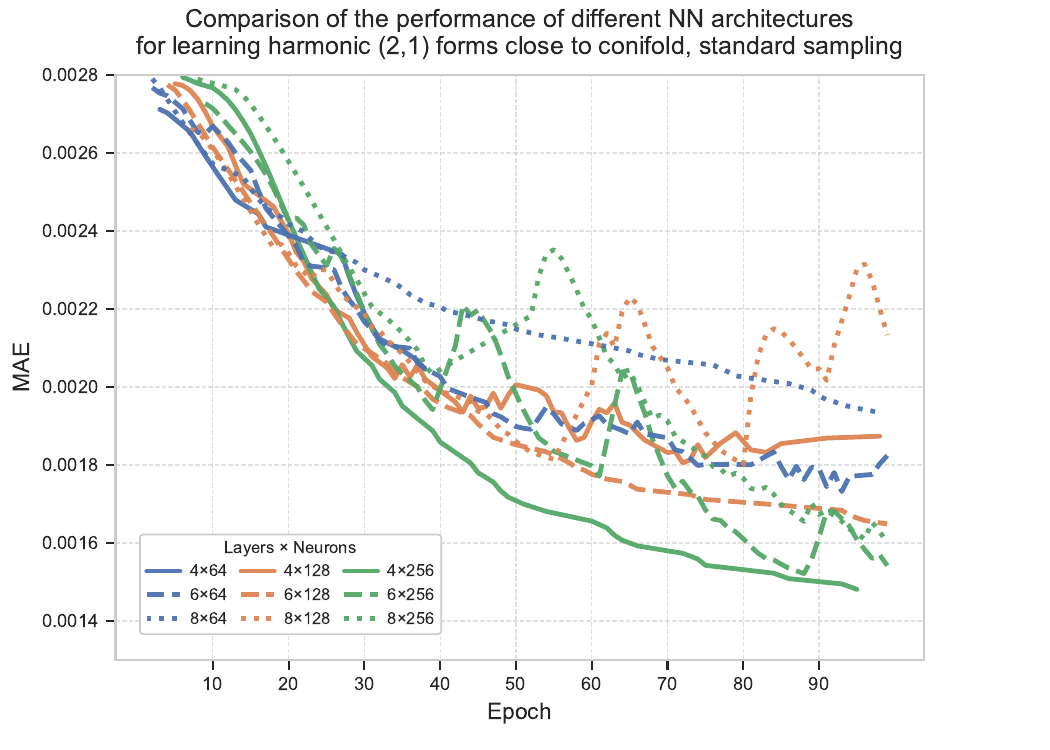} 
    \caption{Comparison of NN architectures for approximating the harmonic representatives using IPS point sampling (left) and standard (right). For standard sampling, we fixed the activation to GeLU.}
    \label{fig:21forms_comparison}
\end{figure}

As discussed in~\eqref{eq:tildem}, we need to learn a correction  $\Delta m$ to make the reference harmonic (2,1)-forms harmonic with respect to the CY metric. For the Dwork quintic with $h^{2,1}=1$, we parameterize the corresponding $\Delta m$ with a NN and try different architectures, see Figure~\ref{fig:21forms_comparison} for the benchmark point close to the conifold.\footnote{Similar results have been obtained for the solution away from the conifold.} We try models with 2 and 4 hidden layers with 64 and 256 neurons each. We also use layer normalization and look at RELU, GELU, SiLU activation functions. Finally, we compare the performance when using the multi-step and single-step $\phi$ NN to approximate the CY metric. We find that larger models perform better. For the activation function, GELU seems to be the best choice. We also see that models with the (more accurate) twop-step method perform better. Overall, we get more than one order of magnitude improvement between the beginning and the end of training. The results for learning the (2,1)-forms away from the conifold locus are similar. We also tried networks with up to 8 layers with 512 neurons per layer, and ResNets to avoid vanishing gradients. Neither led to better performance, indicating that the limit is the number of points and not NN capacity.

\subsection{Learning the warp factor}
\begin{figure}[t]
    \centering 
    \includegraphics[width=0.6\textwidth]{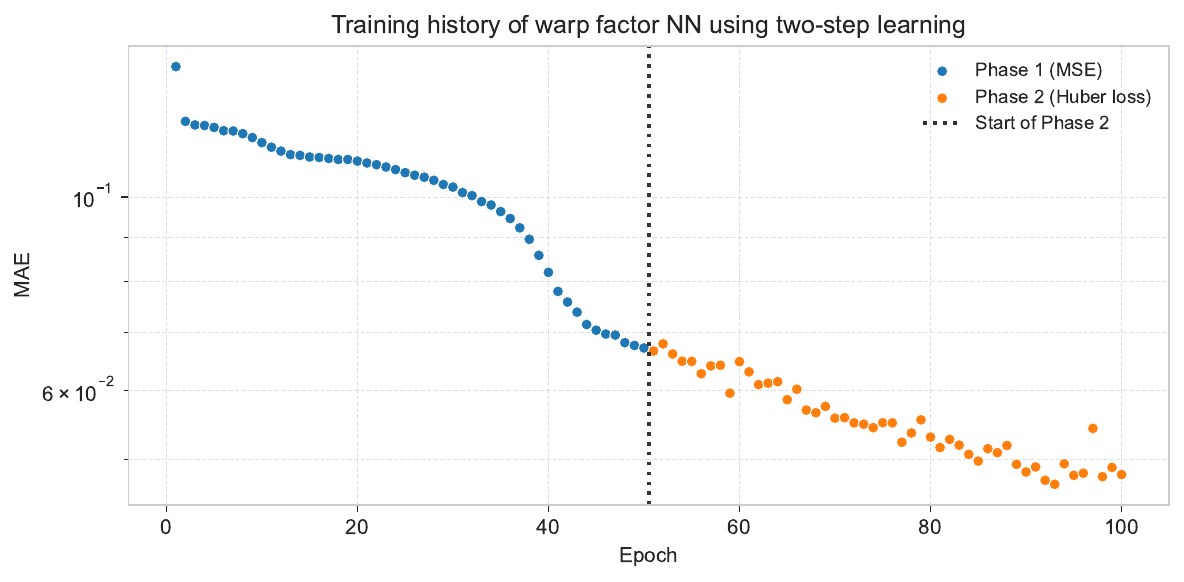} 
    \caption{50 epochs of training the warp factor PINN (4 hidden layers, 256 neurons) with MSE loss, followed by 50 epochs of weighted Huber loss.
    }
    \label{fig:training_weighted_Huber}
\end{figure}

We set up a third NN to learn the warp factor using the two methods outlined in Section~\ref{sec:WarpFactorTheory}. Our architecture hyperparameter scan involves again choosing two or four hidden layer with 64 or 256 neurons each; again, larger architectures did not improve  performance. We show a typical training run in Figure~\ref{fig:training_weighted_Huber}, where the loss drops by around an order of magnitude over the course of the training. Unfortunately, we cannot really benchmark the quality of the warp factor solution other than by noting a drop in the error of the Poisson equation~\eqref{eq:warp_factor}.

\subsection{Size of the throat and the singular bulk problem}

The singular bulk problem~\cite{Gao:2020xqh} states that in KKLT, a large region of the bulk becomes strongly warped, calling into question the control over these solutions. In this section, we want to quantify how ``large'' the region is in which the geometry becomes strongly warped. We discuss our results for different point samples and for training with and without the weighted Huber loss. As mentioned in the Introduction, we study a point in CY moduli space that does not allow for an orientifold involution. We hope to return to a full-fledged string model in a follow-up paper.

To compare the distribution of the warp factor in the CY for different point samples, NNs, losses, and methods of computation, cf.~\eqref{eq:fluxterm1} vs~\eqref{eq:fluxterm2}, we first note that the second-order PDE allows to shift the warp factor $\mathrm{e}^{-4A}$ by a constant corresponding to the overall volume of the warped geometry~\cite{Frey:2008xw}. 
This shift implies that the mean of the warp factor distribution is not fixed, and we will simply align the means of the various distributions we want to compare. A standard way of comparing two probability distributions $P$ and $Q$ is via their KL divergence,
\begin{align}
    D_{\mathrm{KL}}(P \| Q)
    = \int_{\mathbb{R}} 
    P(x)\,\log\!\left(\frac{P(x)}{Q(x)}\right)\, \dd x.    
\end{align}
To compute the KL divergence, we compute the warp factor for many points on the CY and use the histogram for the distributions of $P$ and $Q$. A downside of this is that the result is bin-dependent and sensitive to empty bins in general. This can be ameliorated by using the Kernel-Density-Estimator-based (KDE-based) KL divergence, defined as 
\begin{align}
    D_{\mathrm{KL}}(\hat{P}_h \| \hat{Q}_h)
    =\int\hat{P}_h(x)\log\!\left(
    \frac{\hat{P}_h(x)}{\hat{Q}_h(x)}\right)\, \dd x\,,   
\end{align}
where
\begin{align}
    \hat{P}_h(x)=\frac{1}{n h \sqrt{2\pi}}\sum_{j=1}^{n}\exp\!\left(-\frac{(x-x_j)^2}{2h^2}\right)\,,
\end{align}
is the Gaussian kernel density parameter for the samples $x_i$ and $h$ controls the smoothness. So numerically,
\begin{align}
    D_{\mathrm{KL}}(\hat{P}_h \| \hat{Q}_h)\approx\sum_{k}\hat{P}_h(x_k)\log\!\left(\frac{\hat{P}_h(x_k)}{\hat{Q}_h(x_k)}\right)\Delta x\,.
\end{align}
This is less bin-dependent than the standard KL divergence. Note that the KL divergence is zero for identical distributions and $\infty$ for distributions of disjoint support. We take a KL divergence threshold of $\mathcal{O}(10^{-2})$ to indicate nearly-identical distributions.

To quantify the tail of the distribution, we use the  Complementary Cumulative Distribution function (CCDF), which gives  the probability that a stochastic variable takes a value larger than $x$. Plotting the CCDF, the $y$-axis shows the percentage of points where the warp factor is larger than the value on the $x$-axis. If there are no long tails, this means that the CCDF will drop of sharply around some $x_{\text{max}}$.

Our analysis shows that effects pertaining to the throat are captured poorly if the NN is too small. Even for larger NNs, teasing out the fine-prints of the throat and the warping effects requires using the weighted Huber loss. Using IPS sampling also helps with resolving the throat geometry, since it produces a more uniform sample of the CY geometry with respect to the CY metric, which enters in the PDE for the warp factor.

\subsubsection{Away from conifold}
Away from the conifold, there is no big difference between which point sample we use, see Table~\ref{tab:ConsistencyChecks}, and we will present results for the standard sampling methods. The warp factor distribution between the two methods of solving the warp factor equation (fixing the NN size and the point sampling algorithm) is almost identical, with a KL of 0.025 (for 500 bins) and a KDE-based KL of 0.021. Similarly, changing the NN architectures (we compared two and four layers with 64 and 256 neurons each) does not make a big difference. The biggest effect on the distribution is whether or not we use the Huber loss, but its overall effect is also rather small.

\begin{figure}[t]
\centering
\includegraphics[width=.48\textwidth]{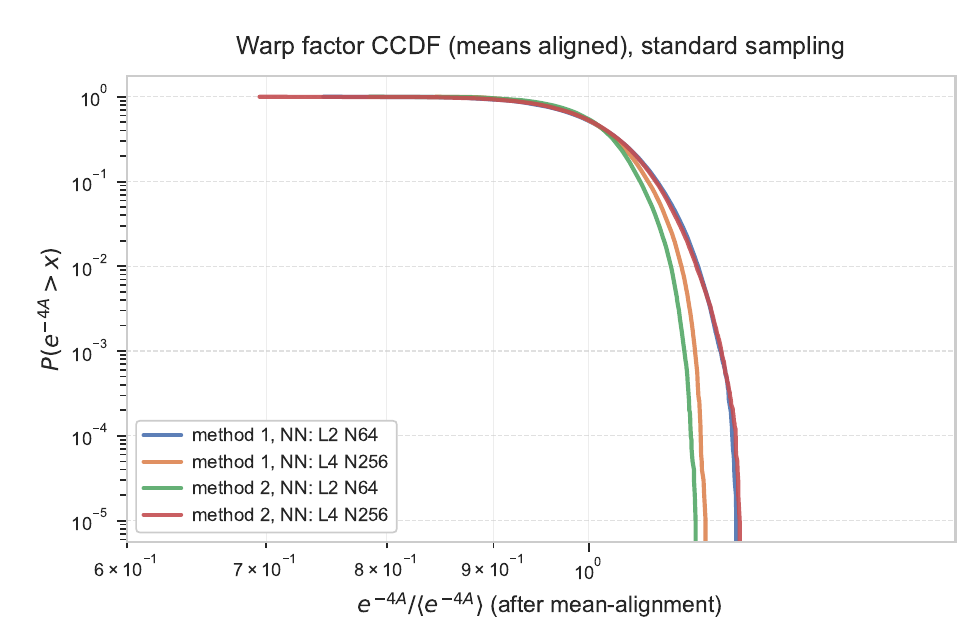}
\includegraphics[width=.48\textwidth]{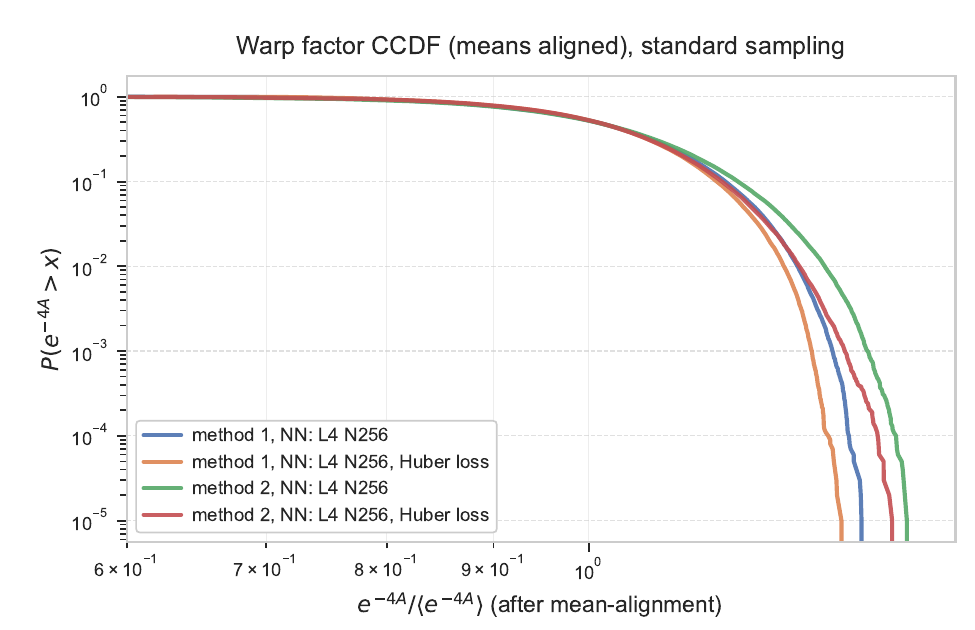}
\caption{Comparison of the CCDFs without (left) and with (right) weighted Huber loss training away from the conifold.}
\label{fig:CCDF_away}
\end{figure}

We also compute the CCDF with and without multi-step metric training. Since we are away from the conifold, we do not expect a strongly warped region, and hence no signal in the CCDF. We use this as a cross-check that the weighted Huber loss is not introducing numerical artifacts by weighing some points much more strongly than others. We present the CCDF for two different NNs and the two different methods without Huber loss on the left, and with Huber loss on the right of Figure~\ref{fig:CCDF_away}. As expected, since there is no throat in this geometry, we see a sharp drop-off of the CCDF, which is also not changed by the inclusion of the Huber loss. With this, we move on to the warp factor close to the conifold region.

\subsubsection{Close to conifold}

\begin{figure}[H]
\centering

\begin{minipage}[t]{0.49\textwidth}
    \centering
    \includegraphics[width=\linewidth]{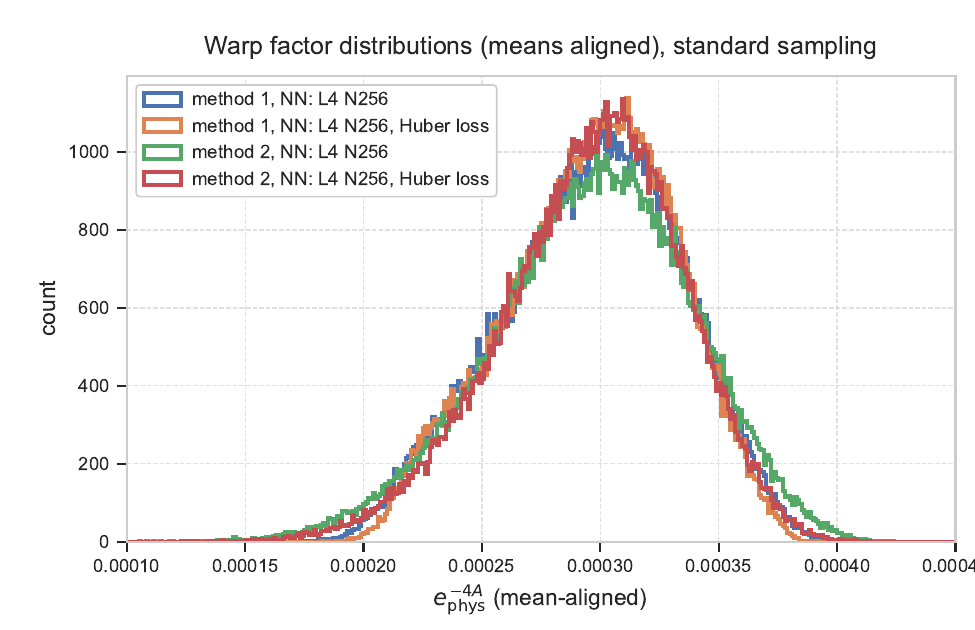}
\end{minipage}
\hfill
\begin{minipage}[t]{0.49\textwidth}
    \centering
    \includegraphics[width=\linewidth]{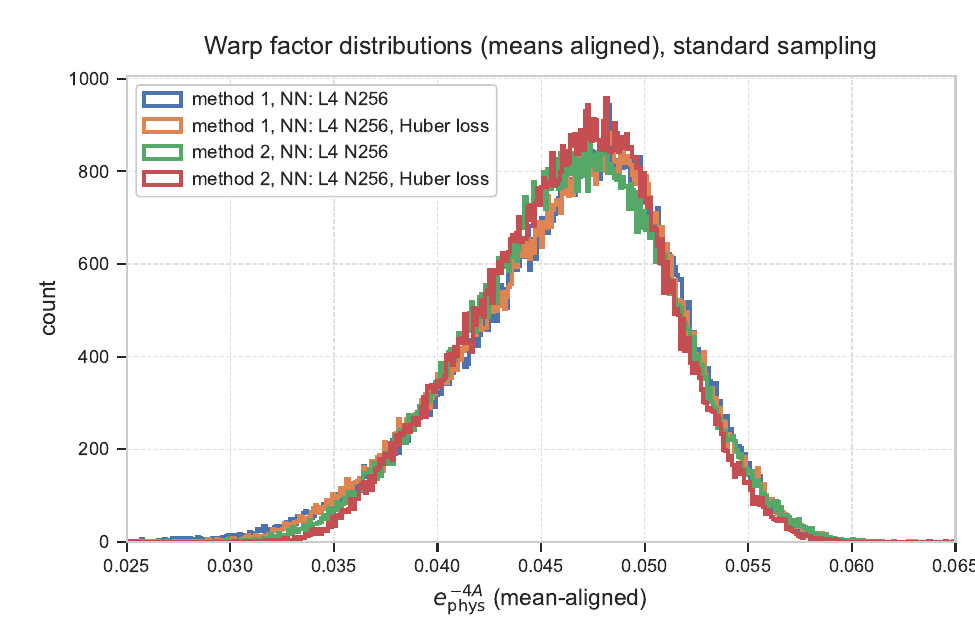}
\end{minipage}

\caption{Comparison of two warp-factor distributions using two-step metric training for our solutions away from and close to the conifold (dataset 4 and 8, respectively). The means of the distributions are aligned.}
\label{fig:warp_dist_away}
\end{figure}

\begin{figure}[t]
\centering
\includegraphics[width=.48\textwidth]{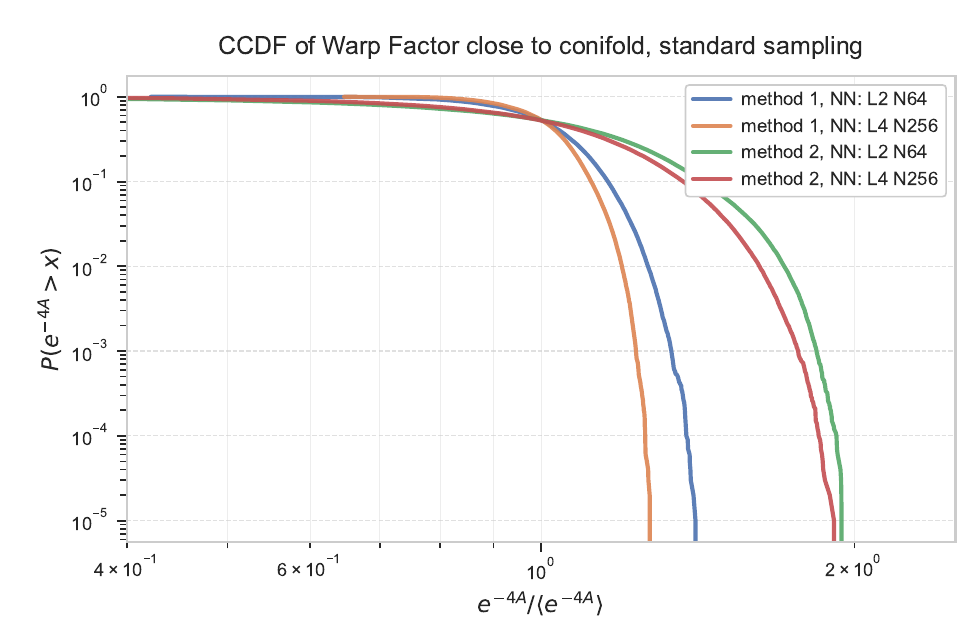}
\includegraphics[width=.48\textwidth]{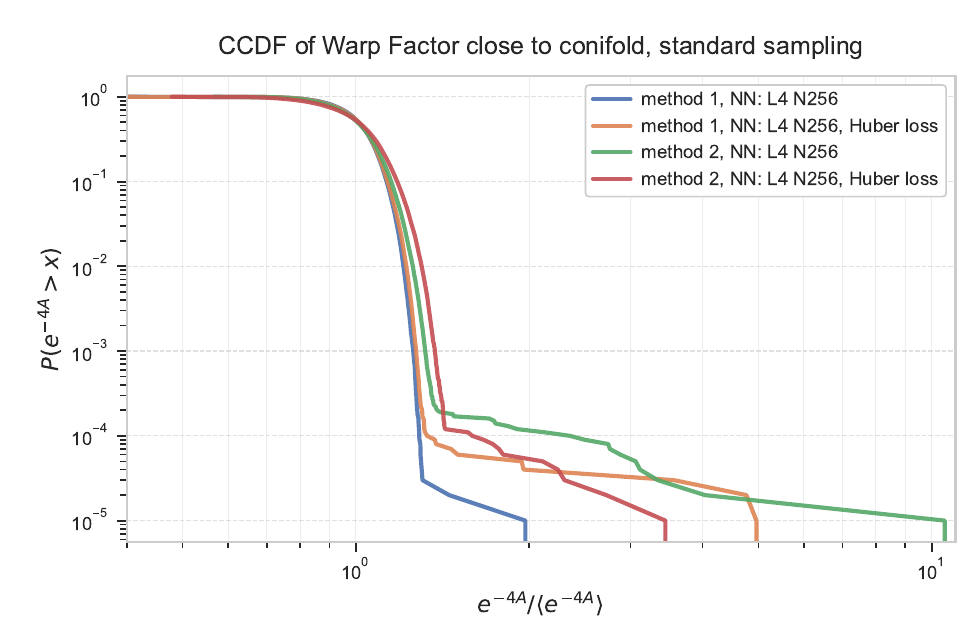}
\caption{Warp factor distribution before and after weighted Huber loss for standard sampling close to the conifold.}
\label{fig:warp_standard_8}
\end{figure}

Before repeating the CCDF analysis close to the conifold, we show the full warp factor distribution we obtain with and without Huber loss in Figure~\ref{fig:warp_standard_8}. We emphasize that the value of the warp factor is only determined up to an integration constant corresponding to the overall volume modulis of the warped compactification.

\begin{figure}[t]
\centering
\includegraphics[width=.48\textwidth]{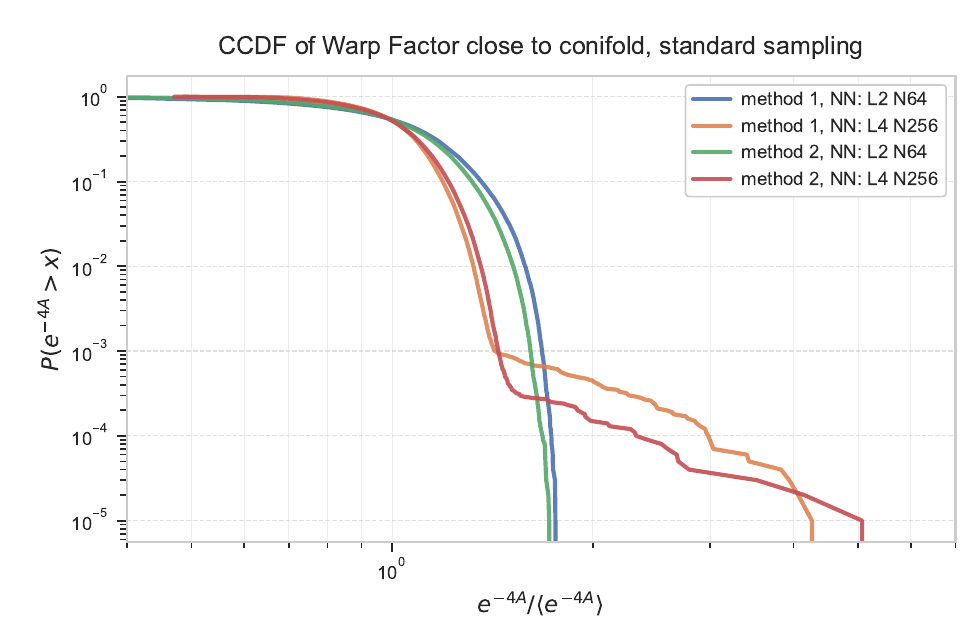}
\includegraphics[width=.48\textwidth]{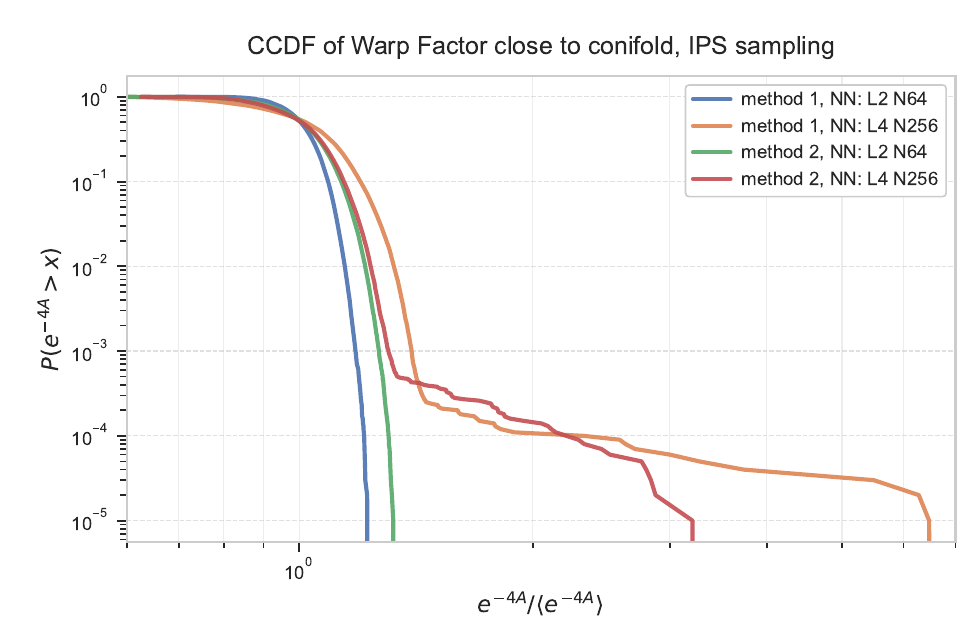}
\caption{Warp factor CCDF for dataset 6 with standard and IPS sampling for different NNs and computation methods, close to the conifold.}
\label{fig:warp_standard_vs_ips}
\end{figure}

First, we do not use Huber loss and compare the result for the standard (left) and IPS (right) point samples in Figure~\ref{fig:warp_standard_vs_ips}. From the comparison we see that the smaller NN does not resolve the throat region (their CCDF looks identical to the one away from the conifold, which has no throat), while the bigger NNs clearly show a different CCDF, independent of whether the warp factor was computed with Method 1 or Method 2. We do see, however, that the throat region is becoming more pronounced in the CCDF for IPS sampling. Again, this is expected, since IPS will also sample regions with strong curvature. This means that those points in the throat that have been sampled with the standard point sampling method should have a relatively large weight in the Monte-Carlo integration. Indeed, we find that for the points indicated to have strong warping by the CCDF, their integration weight is a factor of 5 larger than that of bulk points.

Next, we compare again the KDE-base KL between the two computation methods. For  standard sampling and IPS, we find $0.029$ and $0.15$, respectively. As discussed in Section~\ref{sec:checks}, the variance of quantities computed from IPS-sampled points is much higher, leading to a larger KL-divergence for these datasets.

We repeated this analysis for 20 different point sets with 100k points each (10 point sets sampled with the standard technique and 10 sampled with IP sampling). We find that in some cases, warping in the CCDF is only picked up if we train with weighted Huber loss, see for example Figure~\ref{fig:warp_standard_8}. In cases where the warping is already visible in the CCDF, the weighted Huber loss can help to strengthen the signal.

Finally, we calculate the size of the throat by summing up the weights of the respective points belonging to the bump in the CCDF, which are a few hundred. Dividing this by the sum of all weights yields an estimate for how much of the CY is throat by volume. We find that the throat accounts for approximately $0.5$ percent of the total volume. From the CCDF, we find that a typical point sample has $\mathcal{O}(100)$ points in the throat. But since the Dwork quintic has a discrete symmetry, it actually has $5^3=125$ conifolds and hence the same number of throats. So, on average, we have a single point in each throat. Due to the curse of dimensionality, scaling this up is difficult. One way around it is to modify the IPS point sampler to further zoom in on the strongest curved regions, for example by adjusting $\varepsilon$ in~\eqref{eq:Lambdas}.

Before closing this section, we note that comparing the normalized warp factors away from the conifold and close to it, we see that they are $\mathcal{O}(1)$ and  $\mathcal{O}(10)$, respectively. This means that the integration constant from~\eqref{eq:alphainverseeom}, which is the warped CY volume, has to differ by an order of magnitude as well in order to compare both approximations.

\section{Conclusions and Outlook}
\label{sec:Conclusions}

In this paper, we have constructed a complete numerical pipeline for warped Type IIB flux backgrounds of the GKP type on Calabi-Yau threefolds. Starting from a choice of flux quanta, the pipeline (i) stabilizes the complex structure moduli and the axio-dilaton via the periods, (ii) approximates the Ricci-flat Calabi-Yau metric using the \texttt{cymetric} package, (iii) learns harmonic $(2,1)$-form representatives that encode the imaginary self-dual three-form flux, and (iv) solves the sourced Poisson equation for the warp factor with a physics-informed neural network. We carried out this construction explicitly for the Dwork family of quintics, focusing on two flux vacua: One stabilizing the moduli close to a conifold point, and one near the Landau-Ginzburg point that we used as a numerical cross-check.

Along the way, we introduced several technical improvements that we expect to be of independent interest. We implement the Keller-Lukic sampling strategy~\cite{Keller:2009vj}, which produces samples that are substantially more uniform under the Calabi-Yau measure, in particular in the strongly curved regions near a conifold. We employed a feature-engineered spectral network in the spirit of \cite{Berglund:2022gvm} that automatically respects the projective scaling of the ambient space and removes the need for a transition loss. We showed that multi-step physics-informed training \cite{wang2024multi,AmandaStacked}, in which a frozen first network is refined by a second one, yields roughly an order-of-magnitude improvement in the $\sigma$-measure for the CY metric. We tested second-order optimization with L-BFGS \cite{L-BFGS} for our PINN architectures and concluded that, in the parameter regime relevant for our problem, it does not pay off compared to running a first order optimizer like ADAM for longer. Finally, we introduced a weighted Huber loss for the warp factor PINN that emphasizes rare points in the highly warped throat region and which proved essential for resolving the bump in the warp-factor distribution.

We performed several non-trivial consistency checks of our implementation. The numerical Euler number agrees with the exact value $\chi=-200$ at the sub-percent level away from the conifold and at the $\sim 10\%$ level close to it, with the discrepancy traceable to numerical cancellations in strongly curved regions rather than to errors in the metric network. The Weil-Petersson metric on the complex structure moduli space, computed both from the periods and from the geometry, agrees similarly well. The two computational methods for the warp-factor source, one using the (anti-)self-duality of $\Omega$ and $\chi$, the other explicitly evaluating the Hodge star, yield warp-factor distributions whose KDE-based KL divergence is at the $10^{-2}$ level for standard sampling, indicating excellent agreement.

The main physical application concerns the \emph{singular bulk problem} \cite{Gao:2020xqh} in our toy setup. Resolving the warping in the throat required a combination of larger PINN architectures, IPS sampling, and the weighted Huber loss; with these in place, we observe a clear bump in the complementary cumulative distribution function of $\mathrm{e}^{-4A}$ in the vacuum close to the conifold, while the cross-check vacuum near the Landau-Ginzburg point shows no such feature. Integrating the weights of the points in the bump, we find that the strongly warped region accounts for approximately $0.5\%$ of the total Calabi-Yau volume. The warp factor in this region is an order of magnitude larger as compared to the unwarped geometry, which requires a corresponding suppression by the warped CY volume by the same amount. The Dwork quintic possesses a discrete $\mathbb{Z}_5^3$ symmetry that relates $125$ conifold points, so this volume is naturally distributed over the same number of throats. The price is that a typical sample of $10^5$ points contains only $\mathcal{O}(1)$ point per throat, which sets a hard practical limit on how finely we can resolve the throat geometry with standard sampling.

Several directions are natural to pursue. The most immediate technical issue is the curse of dimensionality just mentioned: Zooming in further on the throat would require a sampler that adaptively refines points in regions of high curvature, e.g.\ by aggressively decreasing $\varepsilon$ in~\eqref{eq:Lambdas} for points already identified as being in the throat, or by combining IPS with importance sampling targeted at the conifold loci. It would also be of interest to study how the volume fraction of the throat region changes as a function of the geodesic distance from the conifold point in complex structure moduli space.
A second direction is to move beyond the Dwork family, which does not admit a consistent orientifold projection~\cite{Carta:2020ohw}, and where we therefore had to mimic the negative tension by a constant smeared source. The natural target is a model from the orientifold landscape of~\cite{Carta:2020ohw} with $h^{2,1}$ small enough to keep the period computation tractable and $h^{1,1}$ small enough for the metric network to converge in reasonable time. In such a setting, the orientifold locus itself sources additional warping, and quantifying its size and overlap with the conifold throat is a quantitative question on which our methods should bear directly. We hope to return to these issues in future work.

\section*{Acknowledgments}
We thank Liam McAllister and Andreas Schachner for useful discussions and comments. The work of FR is supported by the NSF grants PHY-
2210333 and PHY-2019786 (The NSF AI Institute for
Artificial Intelligence and Fundamental Interactions), as well as by startup funding from Northeastern University.

\bibliographystyle{bibstyle}
\bibliography{references}

\end{document}